\documentclass[12pt,a4paper]{article}
\usepackage[utf8]{inputenc}
\usepackage{epsf}
\usepackage{latexsym,amssymb,euscript}
\usepackage[dvips]{graphicx}
\usepackage[numbers,sort&compress]{natbib}
\usepackage{amsmath}
\usepackage{nicefrac}
\usepackage{slashed}
\usepackage{booktabs}
\usepackage[linktocpage]{hyperref}
\usepackage{braket}
\usepackage{chngcntr}
\usepackage{bbold}
\usepackage{multicol}
\usepackage{graphics}
\usepackage{graphicx}
\usepackage{mciteplus}
\usepackage{caption}
\usepackage{subcaption}
\usepackage{pdfpages}
\usepackage[titletoc]{appendix}
\usepackage[normalem]{ulem}
\graphicspath{{./figures/}}
\hypersetup{
 linktocpage = false,
 urlcolor = linkred,
 colorlinks = true,
 linkcolor = linkred,
 anchorcolor = citegreen,
 citecolor = urlblue,
 pdfstartview = {XYZ null null 1.25} 
           }
\usepackage[left=2cm, right=2cm]{geometry}
\usepackage{pstricks}
\usepackage{color}
\usepackage{xcolor}
\definecolor{urlblue}{rgb}{0.2,0.4,0.7}
\definecolor{citegreen}{rgb}{0,0.4,0.2}
\definecolor{linkred}{rgb}{0.9,0.2,0.1}
\usepackage{float}
\usepackage{academicons}
\definecolor{orcidlogocol}{HTML}{A6CE39}
\usepackage{changepage}
\usepackage{fancyhdr}
\newcommand{\drv}{{\rm d}}

\newcommand{\LQCD}{\Lambda_{\rm QCD}}
\newcommand{\MSb}{\overline{\rm MS}}

\newcommand{\LL}{{\rm LL/LO}}

\newcommand{\NLLp}{{\rm NLL/NLO^+}}
\newcommand{\NLLpp}{{\rm NLL/NLO^{(+)}}}
\newcommand{\HENLOp}{{\rm HE}\mbox{-}{\rm NLO^+}}

\newcommand{\ClHENLOp}{{\cal C}_l^{{\rm HE}\text{-}{\rm NLO}^+}}
\newcommand{\CmLL}{{\cal C}_m^\LL}

\newcommand{\CmNLLp}{{\cal C}_m^\NLLp}

\newcommand{\DY}{\Delta Y}

\newcommand{\vqTTa}{\langle {\vec q}_T^{\;2} \rangle}

\newcommand{\E}{{\cal E}}

\newcommand{\Jpsi}{J/\psi}

\newcommand{\BCs}{B_c(^1S_0)}
\newcommand{\Bss}{B_c(^3S_1)}

\newcommand{\XQq}{X_{Qq\bar{Q}\bar{q}}}

\newcommand{\Xcu}{X_{cu\bar{c}\bar{u}}}
\newcommand{\Xcs}{X_{cs\bar{c}\bar{s}}}

\newcommand{\TQQ}{T_{4Q}}
\newcommand{\TQc}{T_{4c}}

\newcommand{\PQQ}{P_{5Q}}
\newcommand{\PQc}{P_{5c}}

\newcommand{\bPQc}{\bar{P}_{5c}}

\newcommand{{\HFNRevo}}{\tt HF-NRevo}

\newcommand{{\Jethad}}{\tt JETHAD}
\newcommand{{\symJethad}}{\tt symJETHAD}
\newcommand{{\psymJethad}}{\tt (sym)JETHAD}
\newcommand{{\Hell}}{\tt HELL}
\newcommand{{\RadISH}}{\tt RadISH}
\newcommand{{\Pegasus}}{\tt QCD-PEGASUS}
\newcommand{{\HOPPET}}{\tt HOPPET}
\newcommand{{\QCDNUM}}{\tt QCDNUM}
\newcommand{{\APFEL}}{\tt APFEL}
\newcommand{{\APFELpp}}{\tt APFEL++}
\newcommand{{\APFELppp}}{\tt APFEL(++)}
\newcommand{{\EKO}}{\tt EKO}
\newcommand{{\FeynCalc}}{\tt FeynCalc}

\newcommand{\eref}[1]{~\eqref{#1}}

\newcommand{\orcidFGC}{\href{https://orcid.org/0000-0003-3299-2203}{\includegraphics[scale=0.1]{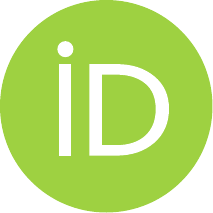}}}

\begin{document}

\begin{titlepage}

{
\begin{adjustwidth}{-1cm}{-1cm}
\begin{center}
  {\LARGE \bf Heavy-flavor multimodal fragmentation \vskip.025cm to $S$-wave pentacharms \vskip.145cm at next-generation hadron colliders}
\end{center}
\end{adjustwidth}
}

\vskip 0.75cm

\centerline{
Francesco~Giovanni~Celiberto$^{\;1\;\dagger}$ {\orcidFGC}
}

\vskip .4cm

\centerline{${}^1$ {\sl Universidad de Alcal\'a (UAH), E-28805 Alcal\'a de Henares, Madrid, Spain}}
\vskip 1.15cm

\begin{abstract}
\vspace{0.25cm}
\headrule 
\vspace{0.50cm}
We investigate the leading-power fragmentation of fully charmed pentaquark states ($S$-wave $|c\bar{c}ccc\rangle$ pentacharms) at hadron colliders.
We introduce a new set of multimodal collinear fragmentation functions, named {\tt PQ5Q1.0}.
They rely on an enhanced calculation of the initial-scale input for the constituent heavy-quark fragmentation channel, making them well suited to describe the short-distance emission of either a compact multicharm state or a dicharm-charm-dicharm configuration.
To explore phenomenological implications, we use the {\tt (sym)JETHAD} multimodular interface to study NLL/NLO$^+$ semi-inclusive production rates for penta\-charm-plus-jet systems at the forthcoming HL-LHC and the future FCC.
Our analysis represents a further step toward bridging the domains of hadronic structure, precision QCD, and exotic matter.
\vspace{0.50cm} 
\headrule
\vspace{0.75cm}
{
 \setlength{\parindent}{0pt}
 \textsc{Keywords}: \vspace{0.15cm} \\ 
 Hadronic structure \\
 High-energy QCD \\
 Exotic matter \\
 Pentacharms \\
 Heavy flavor \\
 Fragmentation \\
 Resummation \\
 {\tt PQ5Q1.0} FF release
\vspace{0.65cm} 
\headrule
}
\end{abstract}

\vspace{-0.00cm}
\vfill
$^{\dagger}${\it e-mail}:
\href{mailto:francesco.celiberto@uah.es}{francesco.celiberto@uah.es}

\end{titlepage}

\tableofcontents
\clearpage

\section{Opening remarks}
\label{sec:intro}

The study of exotic hadrons has been one of the most exciting developments in hadron spectroscopy over the past two decades~\cite{Brambilla:2019esw,Yuan:2018inv,JPAC:2021rxu}. Although the conventional quark model, first proposed by Gell-Mann and Zweig in the nineteenth-sixties, describes baryons as three-quark states and mesons as quark-antiquark pairs~\cite{Gell-Mann:1964ewy,Zweig:1964jf}, recent discoveries suggest a much richer spectrum of hadronic matter. 
The existence of hadrons beyond the traditional classification, such as tetraquarks, pentaquarks, hexaquarks, and hybrid mesons, has fundamentally challenged our understanding of Quantum Chromodynamics (QCD)~\cite{Jaffe:1976ih,Jaffe:1976yi,Rosner:1985yh,Pepin:1998ih,Vijande:2011im,Esposito:2016noz,Lebed:2016hpi,Guo:2017jvc,Lucha:2017mof,Ali:2019roi}. 

These exotic states provide an invaluable opportunity to probe the strong interaction in previously unexplored ways.
Next-generation (lepton-)hadron colliders, such as the LHC Hi-Lumi upgrade (HL-LHC)~\cite{Apollinari:2015wtw}, the Electron-Ion Collider (EIC)~\cite{AbdulKhalek:2021gbh,Khalek:2022bzd,Abir:2023fpo} and the Future Circular Collider (FCC)~\cite{FCC:2018byv,FCC:2018evy,FCC:2018vvp,FCC:2018bvk}, will offer us a faultless chance to examine these exotic particles. 
By generating and analyzing exotic matter events in high-energy hadronic collisions, one can delve into their core structure, including the quark and gluon configurations, as well as their production mechanisms.

Recent progress in precision QCD offers new theoretical perspectives for studying these processes. 
Fixed-order calculations, enhanced by all-order resummations, enable robust calculations of cross sections and distributions, which can be compared with experimental data to deepen our understanding of the dynamics involved in exotic matter production.
A comprehensive program that combines advancements in exotic spectroscopy with the systematic application of precision QCD techniques will provide insights into the role of color confinement and the fundamental properties of multiquark bound states. 
Such studies have the potential to deepen our understanding of QCD and the fundamental nature of subnuclear matter.

Various models have been proposed to describe the internal structure of exotic hadrons. 
One of the most prominent approaches is the compact multiquark model, in which exotic hadrons are tightly bound configurations of multiple quarks and antiquarks~\cite{Jaffe:2003sg,Maiani:2015vwa}. 
In this framework, tetraquarks are interpreted as bound states of two quarks and two antiquarks, while pentaquarks consist of four quarks and one antiquark, and hexaquarks are made of three quarks and three antiquarks.

Another widely accepted approach is the hadronic molecular model, which describes exotic hadrons as loosely bound states of conventional hadrons, such as meson-meson or meson-baryon molecules~\cite{Tornqvist:1993ng,Braaten:2003he,Guo:2013sya,Mutuk:2022ckn,Wang:2014gwa,Guo:2017jvc,Karliner:2015ina,Chen:2016qju,Esposito:2023mxw,Grinstein:2024rcu}. 
The molecular picture naturally explains certain features observed in experiments, such as the narrow widths of some exotic states and their proximity to two-hadron thresholds.

A different perspective is provided by the diquark-antidiquark model, which posits that exotic hadrons are composed of tightly bound diquark and antidiquark pairs~\cite{Maiani:2004vq,tHooft:2008rus,Maiani:2013nmn,Maiani:2014aja,Maiani:2017kyi,Wang:2013exa,Grinstein:2024rcu}. 
This model has been particularly useful in explaining the spectrum of tetraquark states, as it predicts the mass ordering and decay patterns of observed resonances.

An alternative explanation comes from the hadroquarkonium model, in which an exotic hadron consists of a heavy quarkonium core surrounded by a light quark cloud~\cite{Dubynskiy:2008mq,Voloshin:2013dpa,Eides:2019tgv,Ferretti:2018tco,Ferretti:2020ewe}. 
This picture has been applied to states such as the $X(3872)$~\cite{Choi:2003ue} and $Y(4260)$~\cite{CLEO:2006ike} states, which exhibit behavior consistent with a hadroquarkonium-like structure.

The search for exotic hadrons has been a key focus of experimental collaborations worldwide. 
The first evidence for tetraquarks emerged with the discovery of the hidden-charm $X(3872)$ by the Belle Collaboration in 2003~\cite{Choi:2003ue}. 
This state did not align within the conventional charmonium spectrum, thus broadening the perspective toward exotic interpretations. 
Ten years later, the charged charmonium-like state $Z_c(3900)$, a decay product of the anomalous $Y(4260)$, was simultaneously discovered by BES~III~\cite{BESIII:2013ris} and Belle~\cite{Belle:2013yex}, providing further support for the existence of tetraquarks.
More recently, in 2021, the detection of the $X(2900)$ was reported by the LHCb experiment, representing the first observation of an exotic state with open-charm flavor~\cite{LHCb:2020bls}.

Pentaquarks have also been at the center of experimental efforts.
The necessity of including an antiquark in the leading Fock state makes the experimental identification of many pentaquark signatures challenging. 
If the flavor of the antiquark matches that of any other quark in the quintuplet, it cancels out, causing the particle to appear indistinguishable from its three-quark, nonexotic baryon counterpart. 
Consequently, early pentaquark searches focused on particles where such cancellation did not occur.

The initial claims of a pentaquark signal date back to 2003 when the LEPS and DIANA Collaborations reported the observation of the $\Theta^+$ baryon~\cite{LEPS:2003wug,DIANA:2003uet}.
This aligned with the prediction of a pentaquark state with a mass of 1530~${\rm MeV}$, proposed in 1997~\cite{Diakonov:1997mm}.
However, conflicting results from subsequent searches led to significant skepticism~\cite{CLAS:2005koo,Hicks:2004ge,ParticleDataGroup:2008zun,Praszalowicz:2024mji}.

The breakthrough occurred in 2015 when the LHCb Collaboration at CERN provided clear evidence for two pentaquark states, $P_c(4380)^+$ and $P_c(4450)^+$, in the decay of the $\Lambda_b$ baryon~\cite{Aaij:2015tga}.
Both pentaquarks were observed undergoing strong decays into $[J/\psi \,+\, p]$ systems, classifying them as charmonium-like pentaquarks.

That discovery was corroborated in 2019 with the identification of three new pentaquark states, $P_c(4312)^+$, $P_c(4440)^+$, and $P_c(4457)^+$~\cite{Aaij:2019vzc}.
In 2021, the LHCb experiment reported evidence of a new charmonium-like pentaquark with strangeness, specifically the $P_{cs}(4459)$, observed in the $[J/\psi \,+\, \Lambda]$ distribution from $[\Xi_b^- \,\to \, J/\psi  \,+\, \Lambda  \,+\, K^-]$ decays~\cite{LHCb:2020jpq}.  
Subsequently, in 2022, the LHCb announced the observation of the $P_{\psi_s}^{\Lambda}(4338)$ in flavor-untagged $[B^- \to\, J/\psi \,+\, p \,+\, \bar{p}]$ decays~\cite{LHCb:2022ogu}.
For related experimental studies, see, \emph{e.g.}, Refs.~\cite{LHCb:2016lve,LHCb:2021chn}.

Despite remarkable progress in unraveling the mass spectra and decay properties of exotic hadrons since the discovery of the first exotic hadron, their dynamical production mechanisms remain poorly understood.  
To date, only a handful of model-dependent approaches have been explored, including those based on color evaporation~\cite{Maciula:2020wri} and hadron-quark duality~\cite{Karliner:2016zzc,Becchi:2020mjz}.

Lattice QCD simulations and effective field theories have provided further insight, revealing the subtleties of interquark forces and resonance formation~\cite{Francis:2018jyb,Leskovec:2019ioa,Liu:2019tjn,Bicudo:2022cqi,Alexandrou:2024iwi,Prelovsek:2023sta}. 
More recent theoretical work has explored the role of heavy-quark dynamics in pentaquark systems, suggesting that interactions between heavy mesons and baryons play a crucial role in the formation of observed states~\cite{Xiao:2019aya}.
Furthermore, studies have analyzed the interplay of long-range pion exchanges and short-range quark interactions in binding mechanisms, leading to predictions of new pentaquark candidates with different quantum numbers~\cite{Ali:2017jda,Yamaguchi:2017zmn}.

From a theoretical viewpoint, fully heavy tetraquarks and pentaquarks are arguably the most accessible exotic states to investigate. 
With the heavy-quark mass $m_Q$ being significantly larger than the perturbative threshold, a fully heavy tetraquark can be envisioned as a composite system made up of two nonrelativistic charm quarks and two anticharm quarks. Its lowest Fock state, $|QQ\bar{Q}\bar{Q}\rangle$, is free from contributions arising from valence light quarks or dynamical gluons. 
This is strikingly similar to quarkonia, whose dominant state is simply $|Q\bar{Q}\rangle$. 

This similarity suggests that the theoretical methods used for studying quarkonia may also be applicable to heavy tetraquarks. 
Consequently, while charmonia are often described as QCD ``hydrogen atoms''~\cite{Pineda:2011dg}, fully charmed tetraquarks might be viewed as either QCD ``helium-2 atoms'' or ``hydrogen molecules'', depending on the interpretation.
Analogously, a fully heavy pentaquark, $\PQQ$, can be conceived as a composite system made up of four nonrelativistic charm quarks and one anticharm quark, $| Q \bar{Q} Q Q Q \rangle$.
Fully heavy pentaquarks, following this analogy, can be regarded as QCD ``helium-3 atoms''.

The remarkably large cross sections for $X(3872)$ tetraquark production at high transverse momenta, measured in LHC experiments~\cite{CMS:2013fpt,ATLAS:2016kwu,LHCb:2021ten}, offer crucial insights into its production dynamics.  
These observations present a valuable opportunity to refine theoretical models and explore production mechanisms deeply rooted in high-energy precision QCD, such as the leading-power \emph{fragmentation} of a single parton into the observed tetraquark.

At the same time, the increasing complexity in describing the formation of the exotic necessitates a hadron-structure-driven approach.  
To address this, we recently introduced~\cite{Celiberto:2024beg} and delivered~\cite{Celiberto:2024_TQHL11,Celiberto:2024_TQ4Q11} two novel families of zero-mass variable-flavor number-scheme (ZM-VFNS)~\cite{Mele:1990cw,Cacciari:1993mq} fragmentation functions (FFs), named {\tt TQHL1.1} and {\tt TQ4Q1.1}, which respectively depict the collinear fragmentation of a single parton into doubly heavy ($\XQq$) and fully heavy ($\TQQ$) tetraquarks.
These functions combine nonrelativistic model calculations of the parton initial scale inputs~\cite{Suzuki:1977km,Nejad:2021mmp,Feng:2020riv,Bai:2024ezn} with a DGLAP-evolution treatment that consistently account for heavy-quark evolution thresholds~\cite{Celiberto:2024mex,Celiberto:2024bxu,Celiberto:2024rxa}.

More in general, heavy-hadron fragmentation at leading power provides a unique intersection between hadronic structure and precision QCD. 
The presence of one or more heavy quarks in the lowest Fock state has a twofold impact. 
On the one hand, it complicates the theoretical description of energy fragmentation, requiring hadron-structure-driven modeling of the nonperturbative component. 
On the other hand, it calls for the use of perturbative techniques to compute the short-distance fragmentation component. 
Achieving a precise description of heavy-flavor fragmentation thus demands integrating \emph{the best of both worlds}, where hadron-structure explorations and high-precision QCD calculations work in synergy.

In the present work, we will investigate the leading-power fragmentation of fully charmed pentaquark states ($S$-wave $|c\bar{c}ccc\rangle$ pentacharms, or simply $\PQc$) at new-generation hadron colliders.
We will introduce a new set of collinear fragmentation functions, referred to as {\tt PQ5Q1.0} determinations.
They build on an enhanced treatment of the initial-scale input for the (anti)charm fragmentation channel, making them well suited to describe the short-distance emission of either a compact multicharm state or a dicharm-charm-dicharm configuration.

Defining a flexible approach that allows us to incorporate, within the leading power fragmentation, inputs at the initial scale of different nature, is crucial for a complete description of the pentacharm production mechanism.
Indeed, given that quarks possess finite masses and move dynamically in space, even an initially compact state has a nonzero probability of transitioning into a meson-cluster configuration. 
At the quantum level, this transition occurs through fluctuations that manifest as couplings between the compact state and its internal meson subsystems. 
This interaction tends to deform the initially compact structure, evolving it into a more loosely bound object, reminiscent of a diquark- or molecular-like state~\cite{Sazdjian:2022kaf}.

This phenomenon represents a dynamical mechanism whose precise nature requires solving the four- or five-body bound-state problem in the presence of confining forces.  
A general and complete solution to this problem remains elusive.
Therefore, our {\tt PQ5Q1.0} \emph{multimodal} FFs can serve as a useful guidance for future analyses aimed at discriminating between the (relative weight and connections among) distinct exotic formation dynamics.
In this context, we use the term ``multimodal'' to indicate that the initial fragmentation input is modeled through multiple, physically motivated structural components, reflecting different possible internal configurations of the exotic hadron.

For our phenomenological investigation, we will adopt the $\NLLp$ hybrid-factorization scheme, which consistently incorporates the resummation of leading energy-leading logarithms (LL), next-to-leading ones (NLL), and higher-order contributions (NLL$^+$) within the standard collinear framework at the next-to-leading order (NLO).\footnote{The notation $\NLLp$ was first introduced in recent studies of Mueller--Navelet jets~\cite{Celiberto:2022gji}, and later extended to other high-energy processes involving exotic hadrons~\cite{Celiberto:2024mab,Celiberto:2024beg}. It updates traditional high-energy resummation terminology by explicitly indicating both the logarithmic (NLL) and fixed-order (NLO) accuracy, in line with conventions from precision-QCD resummations.} 
Several formulations of hybrid factorization have been introduced in the literature, particularly in the context of small-$x$ and forward production. 
In this work, we adopt a hybrid scheme in which high-energy logarithmic resummation is consistently embedded into the collinear factorization framework through a suitable combination of resummed and fixed-order components. 
This structure, already employed in previous high-energy analyses~\cite{Bolognino:2021mrc,Celiberto:2022dyf}, allows for a consistent treatment of observables characterized by large rapidity intervals, while maintaining full compatibility with collinear inputs.

In what follows, the {\Jethad} numerical interface, complemented by the {\symJethad} symbolic computation plugin~\cite{Celiberto:2020wpk,Celiberto:2022rfj,Celiberto:2023fzz,Celiberto:2024mrq,Celiberto:2024swu}, will be employed to generate predictions for high-energy observables sensitive to pentacharm-plus-jet signatures. 
Our study will span center-of-mass energies ranging from the 14~TeV HL-LHC to the prospective 100~TeV nominal energy of the FCC.

This article is structured as follows. 
In Sec.~\ref{sec:FFs} we present technical details on the way the novel {\tt PQ5Q1.0} collinear FFs for $\PQc$ states are built.
Section~\ref{sec:hybrid_factorization} gives us insight on the hybrid-factorization setup at $\NLLp$.  
In Sec.~\ref{sec:phenomenology} we provide a phenomenological analysis of rapidity and transverse-momentum rates for the semi-inclusive associated production of $\PQc$ plus jet systems. 
Finally, Sec.~\ref{sec:conclusions} comes with conclusions and prospects.

\section{Heavy-flavor fragmentation to $S$-wave $P_{5c}$}
\label{sec:FFs}

In this section we present our strategy to derive the \emph{multimodal} {\tt PQ5Q1.0} FF family, describing the collinear fragmentation of $S$-wave pentacharms from an initial-scale input for the charm channel based either on the direct picture or the scalar-diquark one.

For the sake of completeness, in Sec.~\ref{ssec:FFs-intro} we briefly review main features of heavy-flavor fragmentation, from heavy-light hadrons to quarkonia and exotic hadrons.
Then, moving to the pentacharm case, direct and diquark inputs are given in Secs.~\ref{ssec:FFs-Q-direct} and~\ref{ssec:FFs-Q-diquark}, respectively.
Finally, the energy dependence of the {\tt PQ5Q1.0} functions is presented and discussed in Sec.~\ref{ssec:FFs-PQ5Q10}.

We performed all the symbolic computations required to build the {\tt PQ5Q1.0} set by making use of {\symJethad}, a \textsc{Mathematica}~\cite{Mathematica_V14-2} plugin of {\Jethad}~\cite{Celiberto:2020wpk,Celiberto:2022rfj,Celiberto:2023fzz,Celiberto:2024mrq,Celiberto:2024swu} suited to the symbolic manipulation of analytic formul{\ae} for ha\-dron\-ic structure and precision QCD.

\subsection{Heavy-flavor fragmentation at a glance}
\label{ssec:FFs-intro}

Unlike light hadrons, the fragmentation mechanism that drives the hadronization of heavy-flavored hadrons exhibits an added layer of complexity. 
This arises from the fact that the masses of heavy quarks in their lowest Fock state fall within the perturbative QCD regime. Consequently, while light-hadron FFs have a pure nonperturbative nature, the initial-scale inputs for heavy-hadron FFs can incorporate some perturbative components.

For singly heavy-flavored hadrons, such as $D$, $B$, or $\Lambda_{c,b}$ particles, the initial fragmentation input can be conceptualized as a two-stage process~\cite{Cacciari:1996wr,Cacciari:1993mq,Helenius:2023wkn}. 
In the first stage, a parton $i$, produced in a hard scattering process with large transverse momentum fragments into a heavy quark $Q$. 
Since the QCD running coupling calculated at the mass of the heavy quark is smaller than one, this first step can be calculated perturbatively.
This part, often referred to as the short-distance coefficient (SDC) for the $[i \to Q]$ fragmentation process, unfolds over a shorter timescale than hadronization. 
The first NLO computation of SDCs for singly heavy hadrons was reported in Ref.~\cite{Mele:1990yq,Mele:1990cw}, with subsequent next-to-NLO studies given in Refs.~\cite{Mitov:2006wy,Blumlein:2006rr,Melnikov:2004bm,Mitov:2004du,Biello:2024zti}.

At later timescales, the heavy quark $Q$ hadronizes into a physical hadron. 
This second stage of the fragmentation process is entirely nonperturbative and is typically described using long-distance phenomenological models~\cite{Bowler:1981sb,Peterson:1982ak,Colangelo:1992kh} or effective field theories~\cite{Georgi:1990um,Eichten:1989zv,Neubert:1993mb}.

The final step in assembling a comprehensive ZM-VFNS FF set for heavy-light hadrons involves incorporating energy evolution effects. 
Starting from the nonperturbative initial-scale inputs, numerical techniques are employed to solve the coupled DGLAP evolution equations at the desired perturbative accuracy.

The two-stage initial-scale fragmentation approach developed for singly heavy hadrons can be suitably extended to quarkonia.
Here, the simultaneous presence of a heavy quark $Q$ and its antiquark $\bar{Q}$ in the lowest Fock state $|Q\bar{Q}\rangle$ makes the description of quarkonium fragmentation more involved.
Modern quarkonium theory relies on an effective formalism, known as nonrelativistic QCD (NRQCD)~\cite{Caswell:1985ui,Thacker:1990bm,Bodwin:1994jh} (see Refs.~\cite{Grinstein:1998xb,Kramer:2001hh,Lansberg:2005aw,Lansberg:2019adr} for a pedagogical introduction).

In NRQCD, heavy-quark and antiquark fields are treated as nonrelativistic degrees of freedom, this turning into a factorization between SDCs, which govern the perturbative production of the $(Q\bar{Q})$ pair, and long-distance matrix elements (LDMEs), which portray the nonperturbative dynamics of hadronization.
NRQCD describes a physical quarkonium state as a linear combination of all possible Fock states, systematically ordered through a double expansion in the strong coupling constant $\alpha_s$ and the relative velocity $v_{\cal Q}$ between the constituent heavy quark and the antiquark.

Notably, NRQCD allows for the exploration of the quarkonium production mechanism across both low and moderate to high transverse-momentum ($p_T$) regimes.
The dominant mechanism at low $p_T$ is the \emph{short-distance} formation of the constituent $(Q\bar{Q})$ pair within the hard scattering and its subsequent hadronization, at larger distances, into the physical quarkonium.
Conversely, when $p_T$ grows, the fragmentation of a \emph{single parton} into the observed hadron plus inclusive radiation competes with the short-distance mechanism, eventually becoming the dominant one.
While the short-distance production can be interpreted as a fixed-flavor number-scheme (FFNS, see, \emph{e.g.}, Ref.~\cite{Alekhin:2009ni} for more details) two-parton fragmentation, embodying genuine higher-power corrections~\cite{Kang:2014tta,Boer:2023zit,Celiberto:2024mex,Celiberto:2024bxu,Celiberto:2024rxa}, the single-parton production is a ZM-VFNS fragmentation, its energy evolution being regulated by the DGLAP equations.

Leading-order (LO) calculations of the initial-scale input for gluon and constituent heavy-quark channels into $S$-wave vector quarkonia in color-singlet configurations were conducted in the early nineties~\cite{Braaten:1993rw,Braaten:1993mp},
while corresponding analyses at NLO came out only recently~\cite{Zheng:2019gnb}.
Starting from these inputs, a pioneering determination of new DGLAP-evolving FFs for vector quarkonia, named {\tt ZCW19$^+$}, was derived in Refs.~\cite{Celiberto:2022dyf,Celiberto:2023fzz}.
The {\tt ZCFW22} extension to $\BCs$ and $\Bss$ states soon followed~\cite{Celiberto:2022keu,Celiberto:2024omj}.

Remarkably, the behavior of rapidity and transverse-momentum distributions of those char\-med $B$ me\-sons, determined using the {\tt ZCFW22} FF framework, supported the observation by the LHCb Collaboration~\cite{LHCb:2014iah,Celiberto:2024omj} that the production rate of $\BCs$ mesons relative to singly bottomed $B$ mesons remains below 0.1\%~\cite{Celiberto:2024omj}.
This outcome provided a critical reference point for validating the ZM-VFNS fragmentation scheme at high transverse momentum.

Turning our attention to the exotic sector, recent studies indicate that NRQCD factorization can be leveraged to explore the intrinsic nature of double $\Jpsi$ excitations~\cite{LHCb:2020bwg,ATLAS:2023bft,CMS:2023owd}, interpreting these states as fully charmed compact tetraquarks (tetracharms)~\cite{Zhang:2020hoh}.  
Within this framework, the formation of a $\TQc$ state originates from the short-distance production of two charm and two anticharm quarks, occurring at a scale being approximately the inverse of the charm mass. 
As for singly heavy hadrons and quarkonia, asymptotic freedom enables the theoretical description of heavy-tetraquark fragmentation as a two-step convolution, incorporating a short-distance component followed by a long-distance phase.  

The first calculation of the NRQCD initial-scale input for the $[g \to \TQc]$ $S$-wave fragmentation channel in color-singlet configurations was presented in Ref.~\cite{Feng:2020riv}. 
Then, our recent work on {\tt TQ4Q1.0} FF sets combined the NRQCD gluon channel with an initial-scale input for the $[c \to \TQc]$ FF, derived by suitably adapting a Suzuki-model calculation~\cite{Suzuki:1977km,Nejad:2021mmp}. 
This approach was previously applied to describe the fragmentation of doubly heavy $\XQq$ states~\cite{Nejad:2021mmp}. 

Expanding on this methodology, the first determination of FFs for heavy-light tetraquarks, referred to as {\tt TQHL1.0} functions, was presented in Ref.~\cite{Celiberto:2023rzw} (see Ref.~\cite{Celiberto:2024mrq} for a review).
Then, by publicly releasing the {\tt TQ4Q1.1}~\cite{Celiberto:2024_TQ4Q11} and {\tt TQHL1.1}~\cite{Celiberto:2024_TQHL11} families, we accounted for also modeling the $[Q \to \TQQ]$ initial-scale FF via NRQCD~\cite{Bai:2024ezn}, improved the $\XQq$ fragmentation description, and extended the analysis to bottomonium-like tetraquarks~\cite{Celiberto:2024beg}.

\subsection{Direct fragmentation}
\label{ssec:FFs-Q-direct}

\begin{figure*}[!t]
\centering
\includegraphics[width=0.475\textwidth]{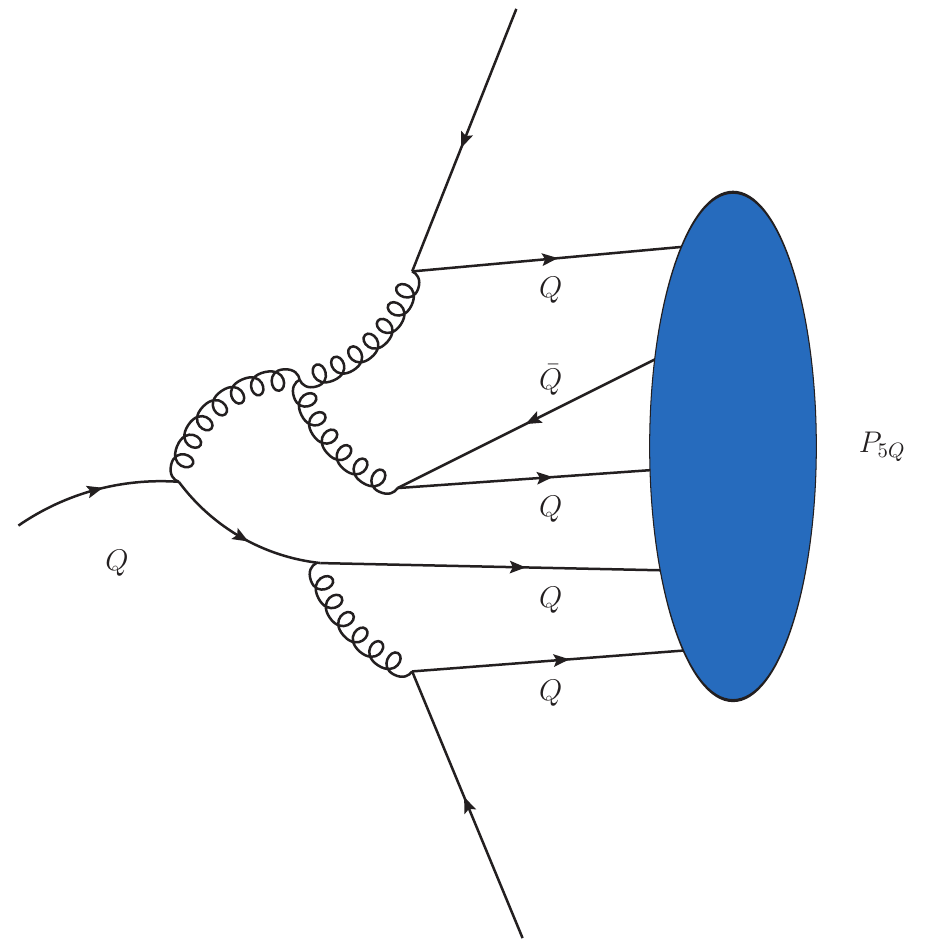}
\hspace{0.40cm}
\includegraphics[width=0.475\textwidth]{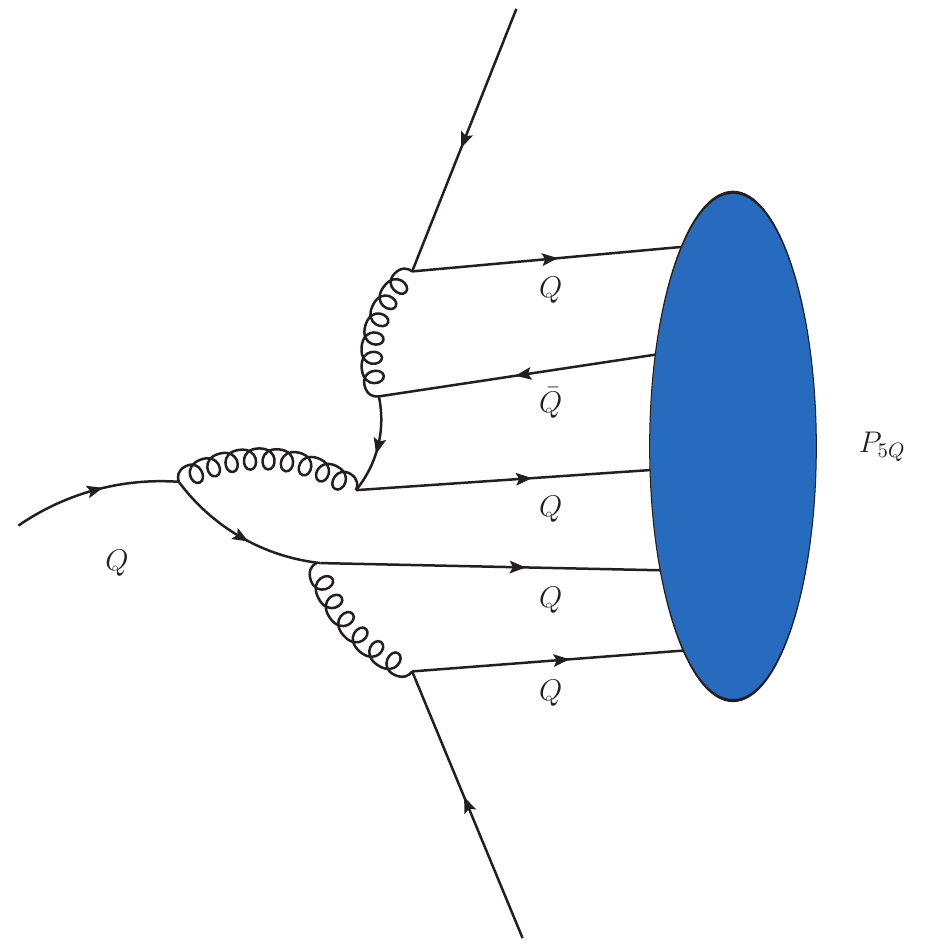}

\vspace{0.20cm}

\includegraphics[width=0.475\textwidth]{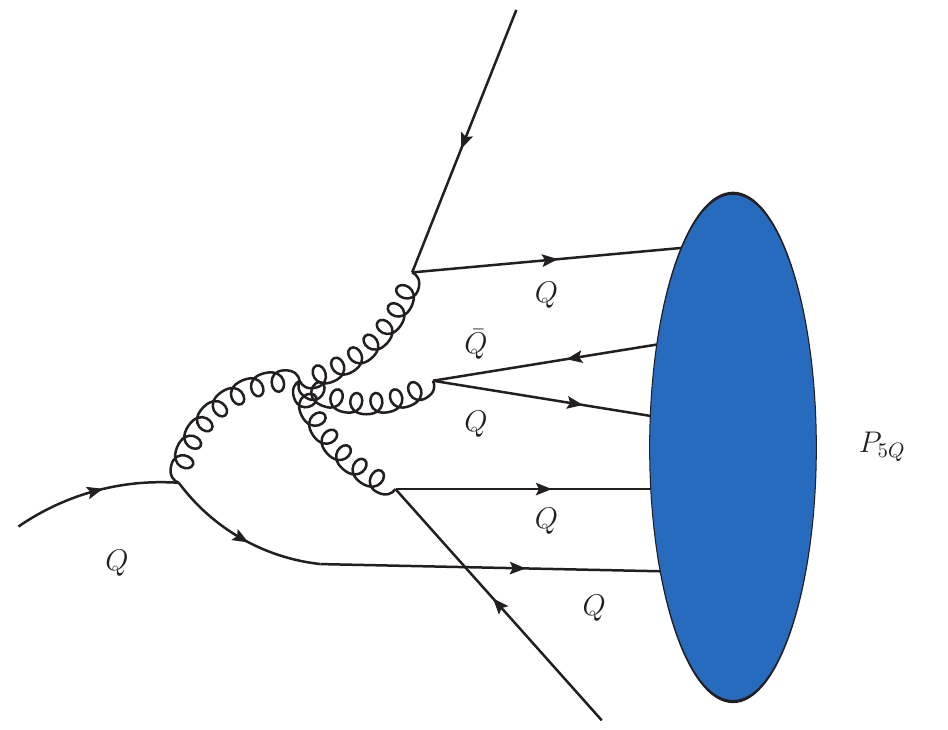}

\caption{Representative leading diagrams for the initial-scale collinear fragmentation of a constituent heavy antiquark into a color-singlet $S$-wave $\PQQ$ pentaquark in the \emph{direct multiquark} picture. 
Blue ovals portray the nonperturbative hadronization component of corresponding FFs. 
Diagrams made with {\tt JaxoDraw 2.0}~\cite{Binosi:2008ig}.}
\label{fig:PQc_FF_direct}
\end{figure*}

Our methodology for modeling the initial-scale input of charm-quark fragmentation into a $S$-wave color-singlet pentacharm in the \emph{direct multicharm} scenario (see Fig.~\ref{fig:PQc_FF_direct}) is based on work done in Ref.~\cite{Farashaeian:2024son}. 
In that study, a spin-physics-inspired Suzuki framework~\cite{Suzuki:1977km}, which incorporates transverse-momentum dependence, was employed. 
The collinear limit was achieved by neglecting the relative motion of the constituent quarks within the bound state~\cite{Lepage:1980fj}.

The proposed approach mirrors the factorization structure of NRQCD fragmentation, where the $(Q\bar{Q})$ pair is perturbatively produced, and the subsequent hadronization is described using the corresponding LDMEs. 
Analogously, in this framework, via a sequence of perturbative splittings above the invariant-mass threshold required to produce all charm quarks present in the final bound state, as depicted in Fig.~\ref{fig:PQc_FF_direct}. 
The production amplitude is then convoluted with a bound-state wave function to describe the nonperturbative tetraquark hadronization dynamics, in accordance with the Suzuki prescription.

We remark that the representative channels shown in Fig.~\ref{fig:PQc_FF_direct} refer to the $[c \to (c \bar{c} c c c) + \bar{c} \bar{c}]$ subprocess, namely the direct $[c + \PQc]$ SDC.
Inverting all charms with anticharms and \emph{vice versa}, one would get the $[\bar{c} \to (\bar{c} c \bar{c} \bar{c} \bar{c}) + \bar{c} \bar{c}]$ splitting, namely the direct $[\bar{c} \to \bPQc]$ SDC.
The analysis of potential imbalances in the production of pentacharms and antipentacharms will be addressed in future dedicated studies. 

In this work, we assume complete symmetry in the formation mechanism of $\PQc$ states and their antiparticles, $\bPQc$, or, equivalently, symmetry in their production rates, considering phenomenological observables sensitive to the inclusive, averaged emission of pentacharms and antipentacharms.
Under this assumption, one has complete symmetry between $c$ and $\bar{c}$ fragmentation channels (for a comparison with the light-hadron case, see, for instance, Ref.~\cite{Bertone:2018ecm}). 

By working with {\symJethad}~\cite{Celiberto:2020wpk,Celiberto:2022rfj,Celiberto:2023fzz,Celiberto:2024mrq,Celiberto:2024swu} as interfaced with {\FeynCalc}~\cite{Mertig:1990an}, we obtained the explicit form of the $[c,\bar{c} \to \PQc]$ initial-scale {\tt PQ5Q1.0} FF in the direct scenario (not given in Ref.~\cite{Farashaeian:2024son}).
It reads
\begin{equation}
\begin{split}
 \label{PQc_FF_initial-scale_Q_direct}
 D^{\PQc}_{c,\,{\rm [direct]}}(z,\mu_{F,0}) \,=\,
 {\cal N}_{P,\,{\rm [direct]}}^{(c)} \,
 (1-z)^4 z^4
 \,
 {\cal R}_{P/c}^2
 \,
 \frac{{\cal S}_{P,\,{\rm [direct]}}^{(c)}(z; {\cal R}_{q_T/c}) }{{\cal T}_{P,\,{\rm [direct]}}^{(c)}(z; {\cal R}_{q_T/c}, {\cal R}_{P/c})}
 \;.
\end{split}
\end{equation}
where we have defined ${\cal R}_{P/c} = M_{\PQc}/m_c$ and ${\cal R}_{q_T/c} = \sqrt{\vqTTa}/m_c$, with $m_c = 1.5$~GeV being the charm-quark mass mass.
As a reasonable assumption, suited to exploratory studies, we will set the pentacharm mass to $M_{\PQc} = 5m_c$.
The overall factor in Eq.~\eqref{PQc_FF_initial-scale_Q_direct} is
\begin{equation}
 \label{PQc_FF_initial-scale_Q_N_direct}
 {\cal N}_{P,\,{\rm [direct]}}^{(c)} \, = \,
 \left\{ 320 \sqrt{5} \pi^2 \, f_{\cal B} \, C_F \big[ \alpha_s\big(\mu_{F,0}^{\rm [direct]}\big) \big]^3 \right\}^2
 \,.
\end{equation}
Here, $f_{\cal B} = 0.25$~GeV stands for the hadron decay constant~\cite{ParticleDataGroup:2020ssz} and $C_F = (N_c^2-1)/(2N_c)$ is the Casimir factor for the emission of a gluon from a quark.

Then, the numerator of 
Eq.~\eqref{PQc_FF_initial-scale_Q_direct} reads 
\begin{equation}
\label{PQc_FF_initial-scale_Q_num_direct}
\begin{split}
 {\cal S}_{P,\,{\rm [direct]}}^{(c)}(z; {\cal R}_{q_T/c}) 
\,&=\,
 \sum\limits_{k=0}^9 \, z^{2k} \, \gamma_{P,\,{\rm [direct]}}^{(c)}(z; k) 
 \left({\cal R}_{q_T/c}\right)^{2k}
\;,
\end{split}
\end{equation}
with the $\gamma_{P,\,{\rm [direct]}}^{(c)}(z; k)$ coefficients given in the Appendix~\hyperlink{app:A}{A}.

Finally, the denominator of 
Eq.~\eqref{PQc_FF_initial-scale_Q_direct} can be cast as 
\begin{equation}
\label{PQc_FF_initial-scale_Q_den_direct}
\begin{split}
 {\cal T}_{P,\,{\rm [direct]}}^{(c)}(z; {\cal R}_{q_T/c}, {\cal R}_{P/c}) 
\,&=\, 
 [{\cal R}_{q_T/c}^2 \, z^2 + (5-3 z)^2]^7 \\[0.20cm]
\,&\times\,
 [{\cal R}_{q_T/c}^2 \, 3z^2 - 13z^2-50 z+75]^2 \\[0.20cm]
\,&\times\, 
 [{\cal R}_{q_T/c}^2 \, (z+2) z^2 + z^3-4 z^2-35 z+50]^2 \\[0.20cm]
\,&\times\, 
 [{\cal R}_{q_T/c}^2 (2z-1) + (1-z)({\cal R}_{P/c}^2-5) + 4]^2
  \;.
\end{split}
\end{equation}

Our $D^{\PQc}_{c,\,{\rm [direct]}}(z,\mu_{F,0})$ FF differs from the one originally introduced in Ref.~\cite{Farashaeian:2024son} in two key aspects.  
First, in the previous work, the ${\cal N}_{P,\,{\rm [direct]}}^{(c)}$ constant in Eq.~\eqref{PQc_FF_initial-scale_Q_N_direct} was not explicitly calculated but rather fixed through a given normalization condition.  
Second, the selection of the $\vqTTa$ parameter in Eq.~\eqref{PQc_FF_initial-scale_Q_direct} needs further scrutiny.  

As previously mentioned, the original approach proposed by Suzuki effectively incorporates spin correlations and represents a proxy for transverse-momentum-dependent (TMD) FFs.  
To obtain the collinear limit, instead of integrating over the squared modulus of the transverse momentum of the outgoing charm quark, one can replace it with its average value, $\vqTTa$.  
This treatment renders $\vqTTa$ a free parameter, which must then be determined on the basis of phenomenological criteria. 
As highlighted in Ref.~\cite{GomshiNobary:1994eq}, larger and larger values of $\vqTTa$ gradually move the FF peak toward the low-$z$ range while simultaneously reducing its overall magnitude.

The initial-scale quark FF derived in Ref.~\cite{Farashaeian:2024son} was based on the choice $\vqTTa = 1\mbox{ GeV}^2$, serving as an upper-bound estimate for the average squared transverse momentum. 
In the present study, we introduce a refined selection for the $\vqTTa$ parameter, grounded in a balanced and thoughtful approach that aligns with the exploratory scope of our work. 

This enhancement builds upon a preliminary adjustment proposed in our earlier analysis of the collinear fragmentation of $\TQc$ states~\cite{Celiberto:2024mab}.
In that study, phenomenological insights derived from the fragmentation production of different hadron species in proton collisions were utilized. 
Specifically, it was observed that heavy-quark FFs for both light hadrons~\cite{Celiberto:2016hae,Celiberto:2017ptm,Bolognino:2018oth,Celiberto:2020wpk} and heavy ones~\cite{Celiberto:2021dzy,Celiberto:2021fdp,Celiberto:2022dyf,Celiberto:2022keu} are typically probed at an average longitudinal fraction value consistently above $\langle z \rangle > 0.4$.

Additionally, it was assumed that constituent-quark FFs are of roughly the same order of magnitude as their corresponding gluon FFs.
This assumption draws support from the simplest quarkonium case, the scalar color-singlet $S$-wave charmonium, $\eta_c$, whose production via fragmentation is described by NRQCD. 
Notably, for $z > 0.4$, the LO fragmentation from both the gluon~\cite{Braaten:1993rw} and the charm~\cite{Braaten:1993mp} to $\eta_c$ is of comparable magnitude.
Through numerical analysis, it was determined that for a charm to $\TQc$ initial-scale FF, setting $\vqTTa_{\TQc} \equiv 70\mbox{ GeV}^2$ yields $\langle z \rangle \gtrsim 0.4$. 
This choice also ensures that the charm fragmentation channel remains of the same order as the corresponding gluon channel.

Currently, no calculations are available for the initial-scale FF of gluons into pentacharms. 
Consequently, to determine the value of the parameter $\vqTTa_{\PQc}$, we rely on its correlation with the peak position of the initial FFs for the charm FF.
Analogous to our approach for the parameter $\vqTTa_{\TQQ}$, a numerical scan over the range of $\vqTTa_{\PQc}$ was performed, resulting in the selection of $\vqTTa_{\PQc} \equiv 90 \, \text{GeV}^2$. 
This choice satisfies the relation
\begin{equation}
 \label{eq:vqTTa_PQc}
 \sqrt{\vqTTa_{\PQc}} \,\approx \, \frac{5}{4} \, \sqrt{\vqTTa_{\TQQ}} \;.
\end{equation}

A more profound justification underpins our choice, extending beyond the heuristic rationale of the relation in Eq.~\eqref{eq:vqTTa_PQc}. 
Seminal investigations into heavy-flavor fragmentation~\cite{Suzuki:1977km,Bjorken:1977md} demonstrated that heavy-quark FFs tend to peak in the large-$z$ region, with binding effects scaling proportionally to the heavy-quark mass.
To illustrate this phenomenon, let us consider the fragmentation production of a $D$ meson, characterized by its lowest Fock state $|c{\bar q}\rangle$, momentum $\kappa$, and mass $m$.

In this context, the constituent heavy quark and the light antiquark must possess approximately the same velocity, denoted as $v \equiv v_c \simeq v_q$. 
Accordingly, their momenta can be defined as $\kappa_c \equiv z \kappa = m_c v$ for the charm, and $\kappa_q = \Lambda_q v$ for the light antiquark, where $\Lambda_q$ represents a hadronic mass scale on the order of $\LQCD$. 
Given that $m \approx m_c$ for a $D$ meson, it follows that $m_c v \approx \kappa = \kappa_c + \kappa_q = z m_c v + \Lambda_q v$. 
This leads to the relationship $\langle z \rangle_c \approx 1 - \Lambda_q/m_c$, where the subscript `$c$' indicates the $[c \to D]$ fragmentation channel.

This intuitive picture is inspired by early insights of Suzuki~\cite{Suzuki:1977hw,Suzuki:1977km}, and was later formulated in a compact kinematic form~\cite{Cacciari:2025_HQF_private}, based on the hypothesis that the heavy quark hadronizes without significant momentum loss. 
The resulting toy model is conceptually aligned with other phenomenological approaches to heavy-flavor fragmentation (see, \emph{e.g.}, Refs.~\cite{Bowler:1981sb,Peterson:1982ak,Colangelo:1992kh,Barger:1977ty}), and also connects to scaling arguments derived from heavy-quark symmetry~\cite{Jaffe:1993ie}.

As highlighted in Ref.~\cite{Celiberto:2024mab}, this behavior may not extend to fully heavy-flavored states, such as quarkonia, $\TQQ$, or $\PQc$ states. 
In these cases, the absence of a soft scale arises because the lowest Fock state lacks light constituent quarks. 
For fully heavy exotics, such as tetracharms or pentacharms, the complex interactions among the four or five constituent heavy quarks complicate the prediction of the fragmentation function peak position, making it difficult to determine solely from kinematic considerations. 

For illustrative purposes, we present the $z$-dependence of $[c \to \PQc]$ initial-scale inputs for our {\tt PQ5Q1.0} determinations. 
To estimate uncertainties associated with our functions at their lowest energy value, we employ a procedure similar to that used in our previous study on tetracharms~\cite{Celiberto:2024mab}. 
In that work, we benchmarked the gluon initial-scale input of the {\tt TQ4Q1.0} functions against the original analysis in Ref.~\cite{Feng:2020riv} by performing a simplified, expanded and diagonal DGLAP evolution that included only the gluon-to-gluon time-like splitting kernel, $P_{gg}$ (refer to Sec. 2.2 of~\cite{Celiberto:2024mab} for technical details). 
Concerning our $[c \to \PQc]$ initial-scale {\tt PQ5Q1.0} FF, we implement a simplified, expanded and diagonal DGLAP evolution exclusively relying upon the quark-to-quark time-like splitting kernel, $P_{qq}$.

Left panel of Fig.~\ref{fig:PQc_FF_initial-scale_Q} illustrates the $z$-dependence of the charm FF in the direct case. 
Shaded bands represent variations in the factorization scale, centered at $\mu_{F,0} = \mu_{F,0}^{\rm [direct]} \equiv 7 m_c$ and spanning from $\mu_{F,0}/2$ to $2\mu_{F,0}$. 
As already explained, this value of $\mu_{F,0}$ is designated as the starting scale for the {\tt PQ5Q1.0} charm direct fragmentation mode. 
Our charm initial-scale FF, multiplied by $z$, presents a pronounced peak in the $0.4 < z < 0.5$ window, and a vanishing pattern at both the $(z \to 0)$ and $(z \to 1)$ endpoints.
The presence of a peak in the moderate to large $z$-region aligns with our expectations, discussed before.

\subsection{Diquark fragmentation}
\label{ssec:FFs-Q-diquark}

\begin{figure*}[!t]
\centering
\includegraphics[width=0.475\textwidth]{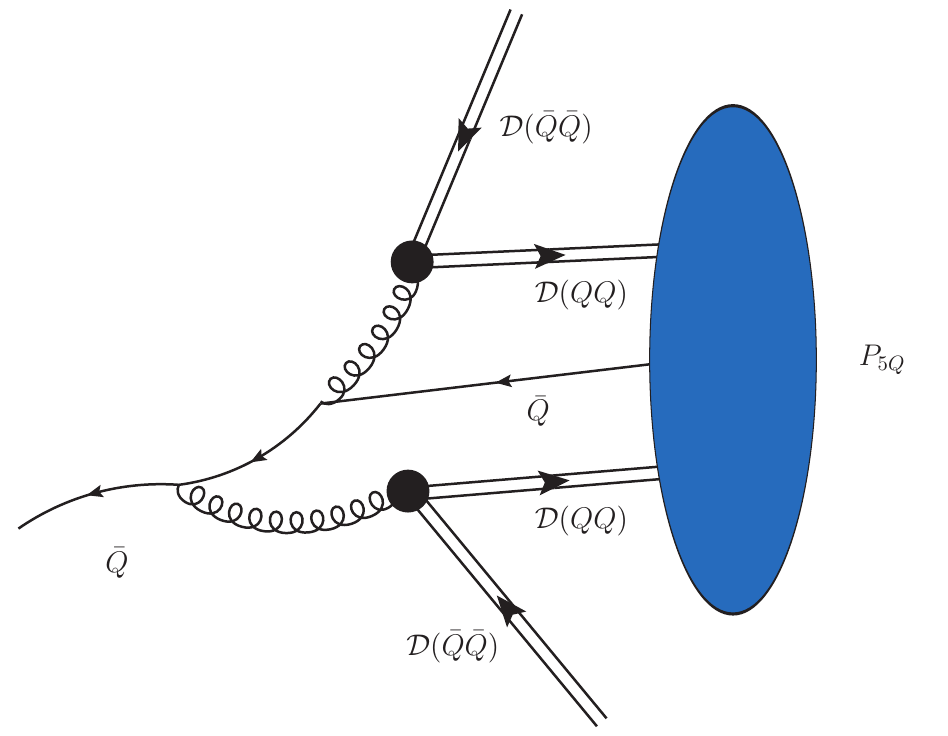}
\hspace{0.40cm}
\includegraphics[width=0.475\textwidth]{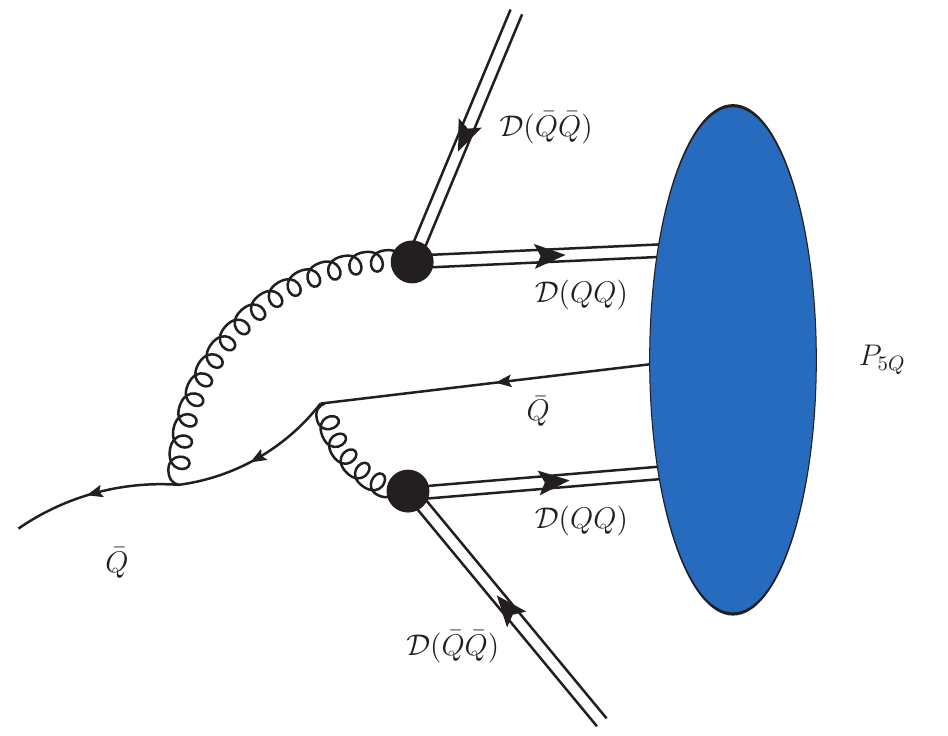}

\vspace{0.20cm}

\includegraphics[width=0.475\textwidth]{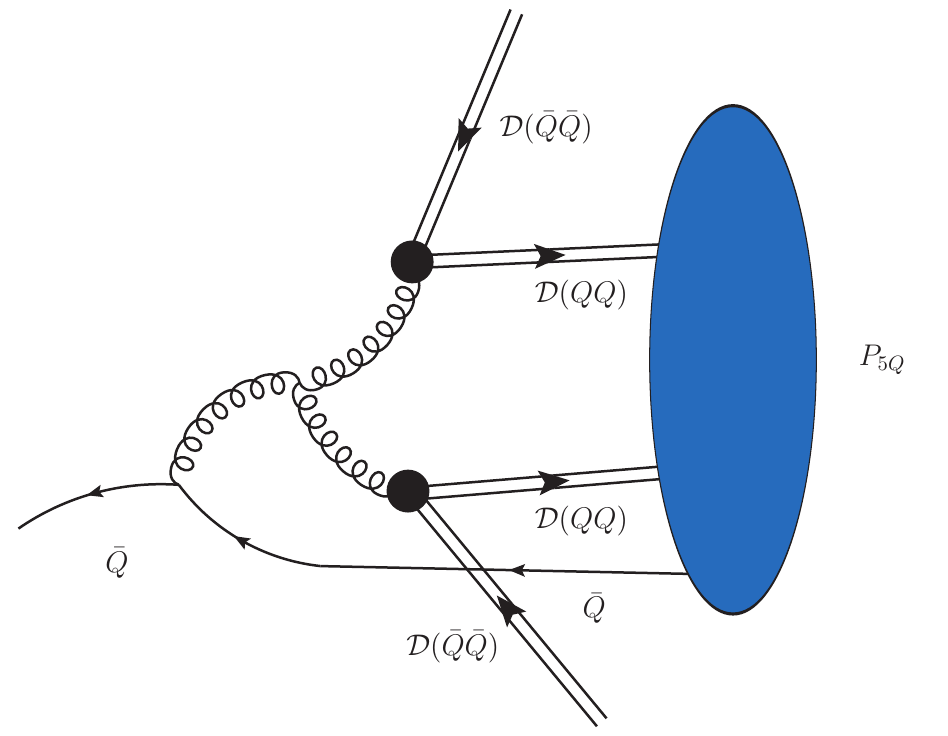}

\caption{Representative leading diagrams for the initial-scale collinear fragmentation of a constituent heavy antiquark into a color-singlet $S$-wave $\PQQ$ pentaquark in the scalar-diquark picture.
Double lines depict for ${\cal D}(QQ)$ or ${\cal D}(\bar{Q}\bar{Q})$ heavy-diquark states, while black bullets are for gluon-diquark-antidiquark effective vertices. 
Blue ovals portray the nonperturbative hadronization component of corresponding FFs. 
Diagrams made with {\tt JaxoDraw 2.0}~\cite{Binosi:2008ig}.}
\label{fig:PQc_FF_diquark}
\end{figure*}

The interacting quark-diquark model posits that two quarks can combine to form a colored quasi-bound state, known as a diquark~\cite{Gell-Mann:1964ewy}. 
This framework has been extensively applied in hadron spectroscopy and extended to the study of baryon production and decay, especially those involving heavy flavors (see Refs.~\cite{Maiani:2004vq,Jaffe:2003sg,Guo:2013xga} for a related discussion). 
In this model, diquarks can exist as either scalar (spin-$0$) or axial-vector (spin-$1$) states.\footnote{Scalar and axial-vector spectator systems are commonly employed in modeling spin-dependent quark~\cite{Bacchetta:2008af} and gluon densities~\cite{Bacchetta:2020vty,Bacchetta:2024fci} within the proton.} 
The composite nature of diquarks is accounted for through phenomenological (nonperturbative) form factors. 

Scalar diquarks require a single form factor, whereas for axial-vector diquarks multiple form factors are needed.
Pioneering applications of the quark-diquark picture to describe the fragmentation production of octet baryons as well as for heavy baryons were presented in Refs.~\cite{Ma:2001ri,Yang:2002gh} and~\cite{Adamov:1997yk,MoosaviNejad:2017rvi,Delpasand:2019xpk}, respectively.
Studies on masses of ground states and resonances of doubly and fully heavy tetraquarks were conducted in Refs.~\cite{Faustov:2020qfm} on the basis of a diquark-antidiquark quasipotential relativistic approach.
Pentaquarks in the diquark picture are extensively discussed in Ref.~\cite{Maiani:2015vwa}.

Our methodology for modeling the initial-scale input of charm-quark fragmentation into a $S$-wave color-singlet pentacharm in the \emph{dicharm-charm-dicharm} scenario (see Fig.~\ref{fig:PQc_FF_diquark}) is based on work done in Ref.~\cite{Farashaeian:2024cpd}. 
As in the direct channel analysis of Sec.~\ref{ssec:FFs-Q-direct}, the spin-physics-inspired Suzuki model~\cite{Suzuki:1977km} is also applied here.
This time, however, the leading Fock state of the pentacharm is assumed to be of the form $| {\cal D}(cc) \, \bar{c} \, {\cal D}(cc) \rangle$, where ${\cal D}(cc)$ stands for a colored heavy diquark composed of two charm quarks.
Within this model, the transition $[\bar{c} \,\to\, ({\cal D}(cc) \, \bar{c} \, {\cal D}(cc) \, \rangle) \,+\, {\cal D}(\bar{c}\bar{c}) \, {\cal D}(\bar{c}\bar{c})]$ is described at LO by the three main classes of Feynman diagrams shown in Fig.~\ref{fig:PQc_FF_diquark}.

We remark that the representative channels shown in Fig.~\ref{fig:PQc_FF_diquark} refer to the diquark-based $[\bar{c} + \PQc]$ SDC.
Inverting all charms with anticharms and \emph{vice versa}, one would get the $[c \,\to\, ({\cal D}(\bar{c}\bar{c}) \, c \, {\cal D}(\bar{c}\bar{c}) \, \rangle) \,+\, {\cal D}(cc) \, {\cal D}(cc)]$ splitting, namely the diquark $[c \to \bPQc]$ SDC.
As already explained in the previous section, we assume complete symmetry in the formation mechanism of pentacharms and their antipentacharms, or, equivalently, symmetry in their production rates.
Thus, as assumed in the direct scenario, also in the diquark one we work with fully symmetry between $c$ and $\bar{c}$ fragmentation channels. 

Treating the pentacharm lowest Fock state as a dicharm-charm-dicharm system yields simpler and more compact analytic expressions compared to the direct multicharm approach.
By making use of {\symJethad}~\cite{Celiberto:2020wpk,Celiberto:2022rfj,Celiberto:2023fzz,Celiberto:2024mrq,Celiberto:2024swu} together with {\FeynCalc}~\cite{Mertig:1990an},
we reproduced the explicit form of the $[c,\bar{c} \to \PQc]$ initial-scale {\tt PQ5Q1.0} FF in the diquark scenario (for technical details, we refer the reader to Sec.~4 of Ref.~\cite{Farashaeian:2024cpd}).
It reads
\begin{equation}
\begin{split}
 \label{PQc_FF_initial-scale_Q_diquark}
 D^{\PQc}_{c,\,{\rm [diquark]}}(z,\mu_{F,0}) \,=\,
 {\cal N}_{P,\,{\rm [diquark]}}^{(c)} \,
 \left[\frac{z^2(1 - z)}{z + 2}\right]^2
 \,
 \frac{{\cal S}_{P,\,{\rm [diquark]}}^{(c)}(z; {\cal R}_{q_T/c}) }{{\cal T}_{P,\,{\rm [diquark]}}^{(c)}(z; {\cal R}_{q_T/c})}
 \;.
\end{split}
\end{equation}
with ${\cal R}_{q_T/c} = \sqrt{\vqTTa}/m_c$\,. 
The overall factor in Eq.~\eqref{PQc_FF_initial-scale_Q_diquark} is
\begin{equation}
 \label{PQc_FF_initial-scale_Q_N_diquark}
 {\cal N}_{P,\,{\rm [diquark]}}^{(c)} \, = \,
 \left[\frac{{\cal V}_{P,\,\rm [diquark]}^{(g{\cal D}\bar{{\cal D}})}}{m_c}\right]^4
 \!
 \left\{ \frac{625 \pi^2}{6 \sqrt{2}} \, f_{\cal B} \, C_F \big[ \alpha_s\big(\mu_{F,0}^{\rm [diquark]}\big) \big]^2 \right\}^2
 \,,
\end{equation}
with ${\cal V}_{P,\,\rm [diquark]}^{(g{\cal D}\bar{{\cal D}})}$ being a form factor entering the description of the gluon-diquark-antidiquark effective vertex (the black bullet in Fig.~\ref{fig:PQc_FF_diquark}). 
Then, the numerator and the denominator respectively read 
\begin{equation}
\label{PQc_FF_initial-scale_Q_num_diquark}
\begin{split}
 {\cal S}_{P,\,{\rm [diquark]}}^{(c)}(z; {\cal R}_{q_T/c}) 
\,&=\,
 \left\{ 64z^4 + 356z^3 - 99z^2 - 2128z + 1540 + \frac{672}{z} \right. \\[0.20cm]
\,&+\, 
 225 \left[ \frac{128z(1 - z)^6}{[{\cal R}_{q_T/c}^2 \, z^2 + (5 - z)^2]^2} - \frac{45(z + 2)^2(8z^2 + 60z - 33 - 60/z)}{[{\cal R}_{q_T/c}^2 \, z^2 + z^2 - 10z + 100]^2} \right] \\[0.20cm]
\,&+\,
 30 \left[ \frac{16(1 - z)^4 (z^2 - 20z - 35)}{{\cal R}_{q_T/c}^2 \, z^2 + (5 - z)^2} \right] \\[0.20cm]
\,&+\,
 \left. \frac{(z + 2)(100z^4 + 256z^3 - 588z^2 + 1859z - 922 - 1380/z)}{{\cal R}_{q_T/c}^2 \, z^2 + z^2 - 10z + 100} \right\}
\end{split}
\end{equation}
and
\begin{equation}
\label{PQc_FF_initial-scale_Q_den_diquark}
\begin{split}
 {\cal T}_{P,\,{\rm [diquark]}}^{(c)}(z; {\cal R}_{q_T/c}) 
  \,&=\, 
 [{\cal R}_{q_T/c}^2 \, z^2 + (5-z)^2]^2
  \;.
\end{split}
\end{equation}

As mentioned, in our study we have taken the calculation presented in Ref.~\cite{Farashaeian:2024cpd} as a proxy reference for the diquark initial-scale charm-to-pentacharm fragmentation.
That function was obtained by considering scalar
diquarks only, and neglecting, for simplicity, pesudovector states.

Our treatment of the $D^{\PQc}_{c,\,{\rm [diquark]}}(z,\mu_{F,0})$ FF enhances the one of Ref.~\cite{Farashaeian:2024cpd} in different aspects. 
First, similar to the direct scenario discussed in the previous section, we do not impose a fixed normalization condition on the overall constant of Eq.~\eqref{PQc_FF_initial-scale_Q_diquark}. 
Instead, as expressed in Eq.~\eqref{PQc_FF_initial-scale_Q_N_diquark}, we explicitly compute it at the initial scale $\mu_{F,0}^{\rm [diquark]} = 9 m_c$ (for further details on the energy evolution of our functions, see Sec.~\ref{ssec:FFs-PQ5Q10} ).
Second, instead of treating $\vqTTa$ as a free parameter, we fix it by following the same approach as in the direct mode (see Eq.~\eqref{eq:vqTTa_PQc}).
Third, to better assess the impact of the genuine scalar-diquark calculation on top of the direct multicharm study, we set the $(g{\cal D}\bar{{\cal D}})$ form factor at ${\cal V}_{P\,,\rm [diquark]}^{(g{\cal D}\bar{{\cal D}})} = 1~\text{GeV}$ instead of $5~\text{GeV}$, as originally proposed in Ref.~\cite{Farashaeian:2024cpd}.

\begin{figure*}[!t]
\centering

\includegraphics[scale=0.645,clip]{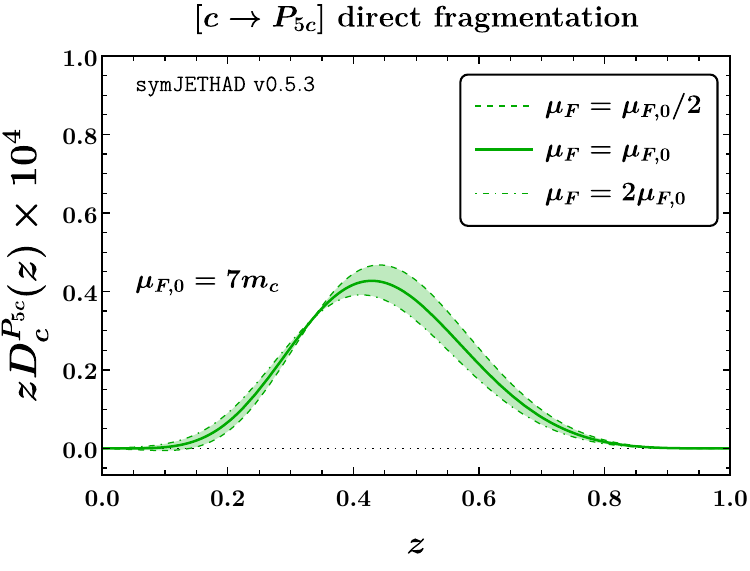}
\hspace{0.25cm}
\includegraphics[scale=0.645,clip]{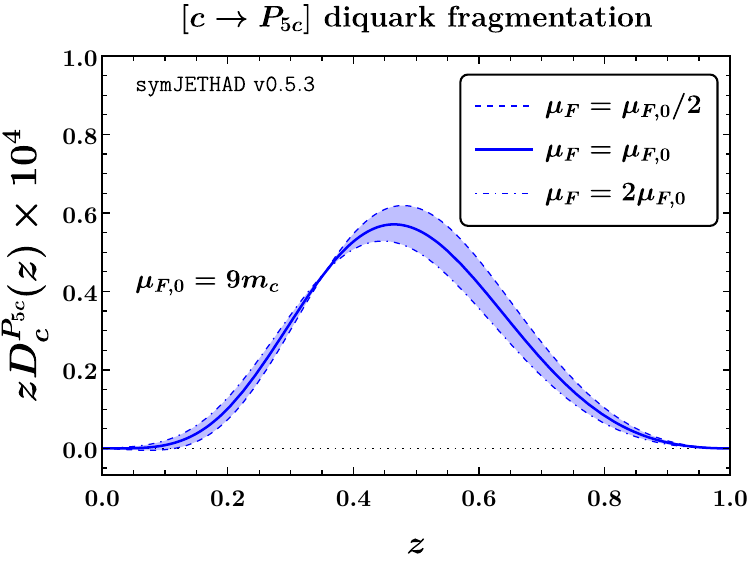
}

\caption{Charm to pentacharm initial-scale fragmentation channel in the direct (left) and scalar-diquark (right) picture. 
For the sake of illustration, an expanded diagonal DGLAP evolution is performed in the range $\mu_{F,0}/2$ to $2\mu_{F,0}$.}
\label{fig:PQc_FF_initial-scale_Q}
\end{figure*}

Right panel of Fig.~\ref{fig:PQc_FF_initial-scale_Q} shows the $z$-dependence of the charm FF in the diquark case. 
Shaded bands represent variations in the factorization scale, centered at $\mu_{F,0} = \mu_{F,0}^{\rm [direct]} \equiv 9 m_c$ and spanning from $\mu_{F,0}/2$ to $2\mu_{F,0}$ via a diagonal DGLAP evolution. 
As anticipated, this value of $\mu_{F,0}$ is designated as the starting scale for the {\tt PQ5Q1.0} charm diquark fragmentation mode. 
In analogy with the direct mode (left panel), our charm initial-scale FF, multiplied by $z$, present a pronounced peak in the $0.4 < z < 0.5$ window, and a vanishing pattern at both the $(z \to 0)$ and $(z \to 1)$ endpoints.
The diquark-FF peak is approximately $10 \div 15 \%$ higher than the direct-FF peak, suggesting higher production rates at high energies in the former case (see Sec.~\ref{sec:phenomenology} for phenomenological applications).

\subsection{The {\tt PQ5Q1.0} functions}
\label{ssec:FFs-PQ5Q10}

The final step in developing our {\tt PQ5Q1.0} FFs for $\PQc$ pentacharms entails applying a consistent DGLAP evolution of the $[c,\bar{c} \to \PQc]$ initial-scale inputs outlined in the previous sections.  
From kinematic considerations, the minimal invariant mass for the $[c \to (c \bar{c} c c c) + \bar{c} \bar{c}]$ splitting in the direct multiquark scenario is $\mu_{F,0}^{\rm [direct]} = 7 m_c$ (see Fig.~\ref{fig:PQc_FF_direct}).
Conversely, the minimal invariant mass for the $[\bar{c} \to (c \bar{c} c c c) + \bar{c} \bar{c} \bar{c} \bar{c}]$ splitting in the scalar-diquark mode is $\mu_{F,0}^{\rm [diquark]} = 9 m_c$ (see Fig.~\ref{fig:PQc_FF_diquark}).
These energy scales are adopted as the fragmentation evolution threshold for the charm quark in each respective mode.

As previously noted, a consistent approach to heavy-hadron fragmentation necessitates integrating our hadronic-structure-driven inputs with collinear factorization, ensuring the accurate determination of DGLAP evolution thresholds for all parton fragmentation channels.
To this end, a novel methodology, known as \emph{heavy-flavor nonrelativistic evolution} ({\HFNRevo}) was defined~\cite{Celiberto:2024mex,Celiberto:2024bxu,Celiberto:2024rxa}.
Designed to depict the DGLAP evolution of heavy-hadron fragmentation from nonrelativistic inputs, and then adapted to more general initial energy-scale models, {\HFNRevo} builds on three key building blocks: \emph{interpretation}, \emph{evolution}, and \emph{uncertainties}.

Focusing on evolution, the DGLAP treatment of {\HFNRevo} FFs can be envisioned as a two-step mechanism.
First, an \emph{expanded} and semi-analytic \emph{decoupled} evolution framework ({\tt EDevo}) ensures a precise treatment of evolution thresholds across all parton channels.
Then, the standard evolution ({\tt AOevo}) is numerically made to \emph{all orders}.

In the context of exotic hadrons, the {\HFNRevo} methodology was initially employed to characterize the collinear fragmentation of fully heavy tetraquarks, utilizing gluon and constituent heavy-quark initial-scale inputs (for a detailed description of $\TQQ$ {\tt TQ4Q1.0} functions and their {\tt 1.1} update, see Ref.~\cite{Celiberto:2024mab} and Sec.~2.3.4 of Ref.~\cite{Celiberto:2024beg}, respectively).

When only one parton FF channel is modeled at the initial energies, a simplified version of {\HFNRevo} is at work.
For instance, the {\tt TQHL1.0} sets (and their {\tt 1.1} update), which depict the fragmentation of doubly heavy $\XQq$ tetraquark states, were constructed by considering only the $[Q,\bar{Q} \to \XQq]$ channel (see Ref.~\cite{Celiberto:2023rzw} and Sec.~2.2.2 of Ref.~\cite{Celiberto:2024beg}).
In such a case, the {\tt EDevo} step is simply skipped and one directly proceeds with the numerical {\tt AOevo} phase.

Coming back to $\PQc$ fragmentation, to the best of our knowledge, the only initial-scale input available to date is the quark channel.
Therefore, similar to the $\XQq$ case, we construct our {\tt PQ5Q1.0} determinations by directly applying the {\tt AOevo} DGLAP evolution to the $[c, c \to \PQc]$ inputs, allowing the other parton channels to emerge exclusively through the evolution process.

In our framework, the charm-quark mass $m_c$ explicitly enters the initial condition for the $[c, \bar{c} \to \PQc]$ channel, through the two-modal, mass-dependent scenario presented in the previous subsections. 
This reflects the physical intuition that fragmentation into a $\PQc$ state takes place near the heavy-quark threshold, where finite-mass effects are relevant.
As a consequence, the DGLAP evolution is consistently performed within a five-flavor scheme, with $n_f = 5$ fixed throughout and no dynamic threshold crossing.

The evolution is carried out in the ZM-VFNS, where all partons, including charm and bottom, are treated as massless. 
This approach is standard in studies of heavy-hadron fragmentation (see, for instance, Ref.~\cite{Cacciari:2024kaa}), and it relies on the fact that mass effects become negligible at the higher scales relevant for the evolution. 
The use of the ZM-VFNS acronym denotes the massless nature of the evolution, in contrast to the mass-dependent modeling adopted at the input scale.

One could argue that our method does not fully incorporate the initial-scale contributions from light partons and nonconstituent heavy quarks, which only become active through evolution at larger energies.
Nevertheless, Ref.~\cite{Nejad:2021mmp} presents reasoning that suggests analogous contributions for FFs of doubly heavy tetraquarks are negligible at the initial scales.
A comparable conclusion is reached for vector-quarkonium FFs in Ref.~\cite{Celiberto:2022dyf}.
We postpone to the future advancements the extension of our treatment by including the remaining parton initial-scale inputs.

\begin{figure*}[!t]
\centering

   \hspace{-0.00cm}
   \includegraphics[scale=0.410,clip]{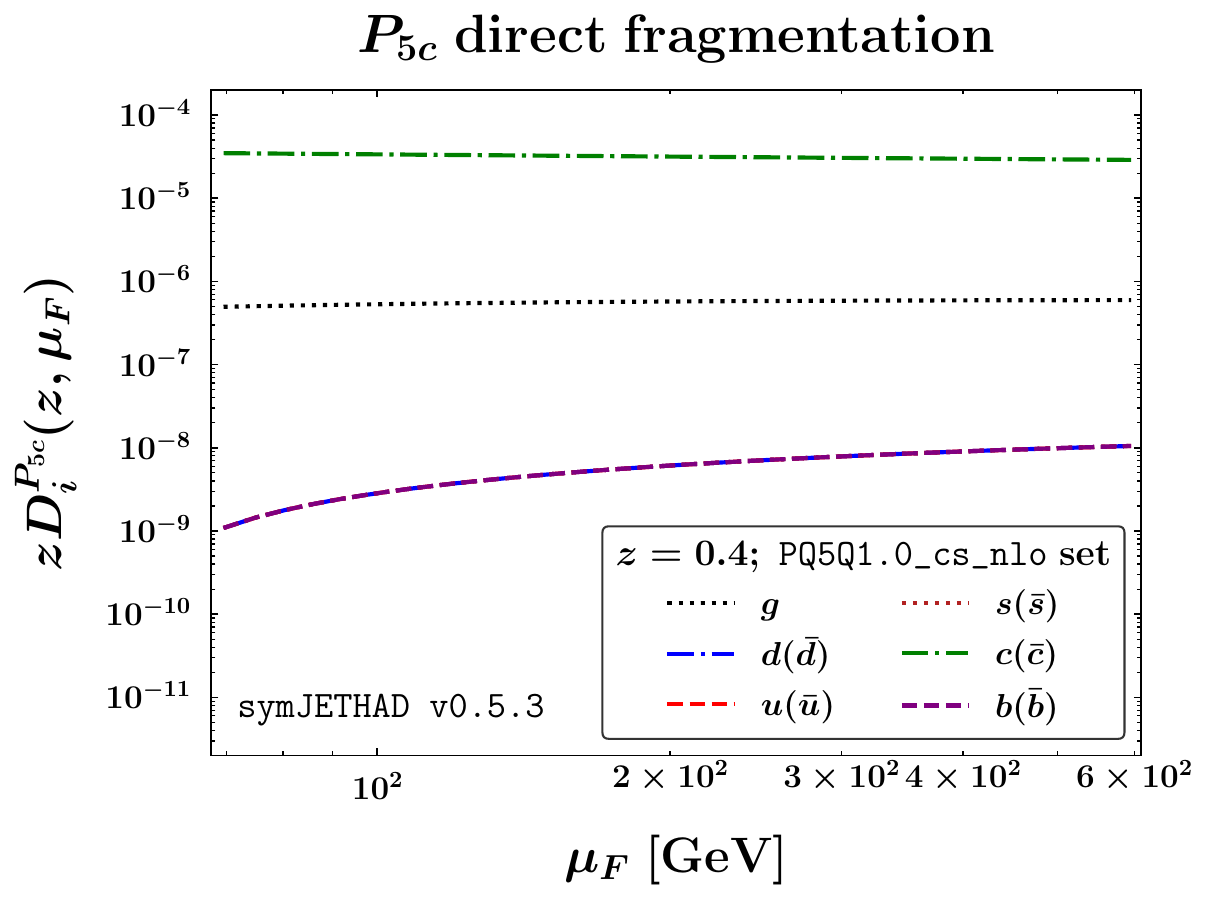}
   \hspace{-0.15cm}
   \includegraphics[scale=0.410,clip]{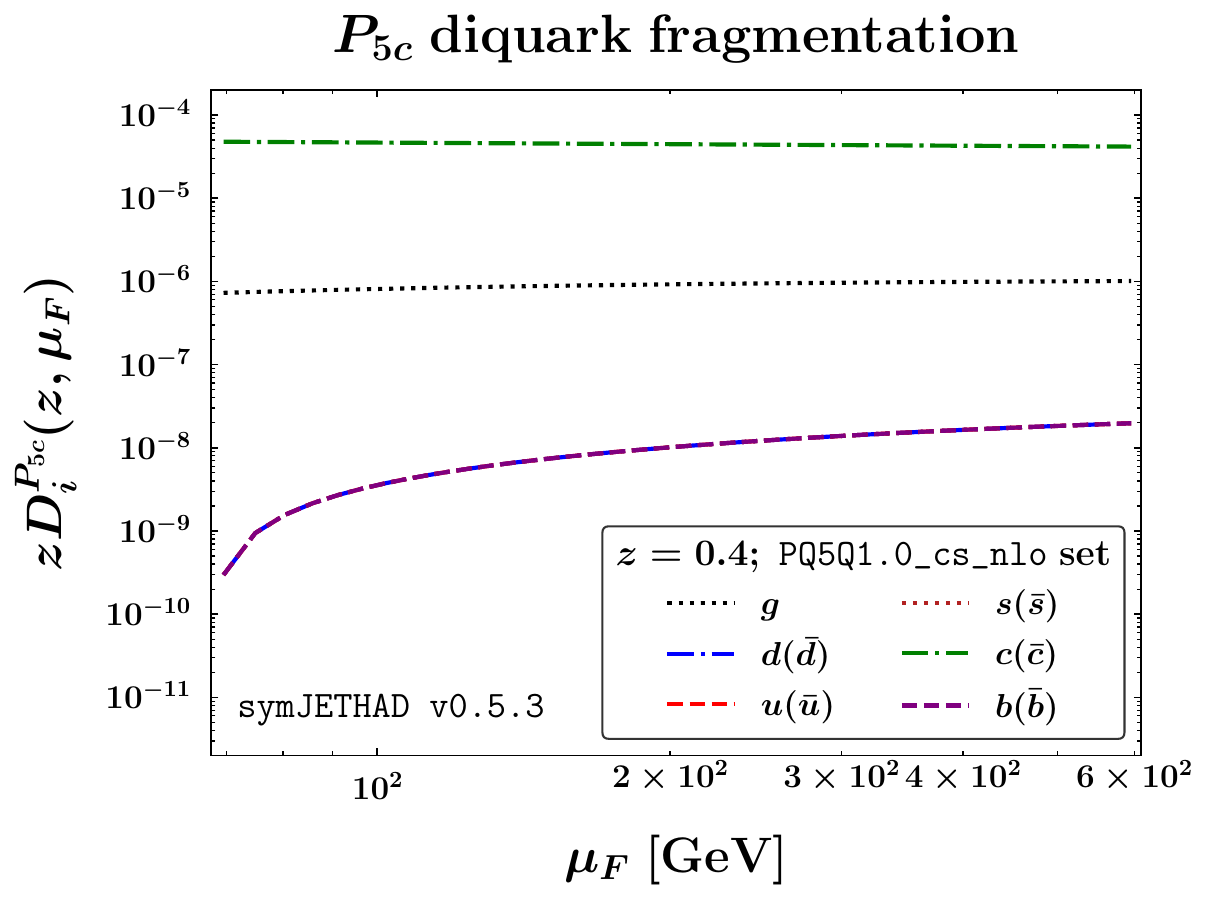}

\caption{Factorization-scale dependence of {\tt PQ5Q1.0} functions describing the collinear fragmentation of $P_{5c}$ states within direct (left) or scalar-diquark (right) initial-scale inputs, at $z = 0.4 \simeq \langle z \rangle$.}
\label{fig:NLO_FFs_Pc0}
\end{figure*}

In Fig.~\ref{fig:NLO_FFs_Pc0}, we show our {\tt PQ5Q1.0} FFs as a function of $\mu_F$ for both direct (left) and scalar-diquark (right) scenarios.  
For simplicity, we focus on a single value of the momentum fraction, $z = 0.4 \simeq \langle z \rangle$, which roughly corresponds to the average value typically probed in semi-hard reactions (see, for instance, Refs.~\cite{Celiberto:2016hae,Celiberto:2017ptm,Celiberto:2020wpk,Celiberto:2021dzy,Celiberto:2021fdp,Celiberto:2022dyf,Celiberto:2022keu,Celiberto:2022kxx,Celiberto:2024omj}).
We note that the charm to pentacharm FF significantly outperforms both the light-parton and nonconstituent heavy-quark channels. 
It consistently remains about one to two orders of magnitude higher than the gluon channel throughout the entire examined energy range. 

\begin{figure*}[!t]
\centering

   \hspace{-0.00cm}
   \includegraphics[scale=0.410,clip]{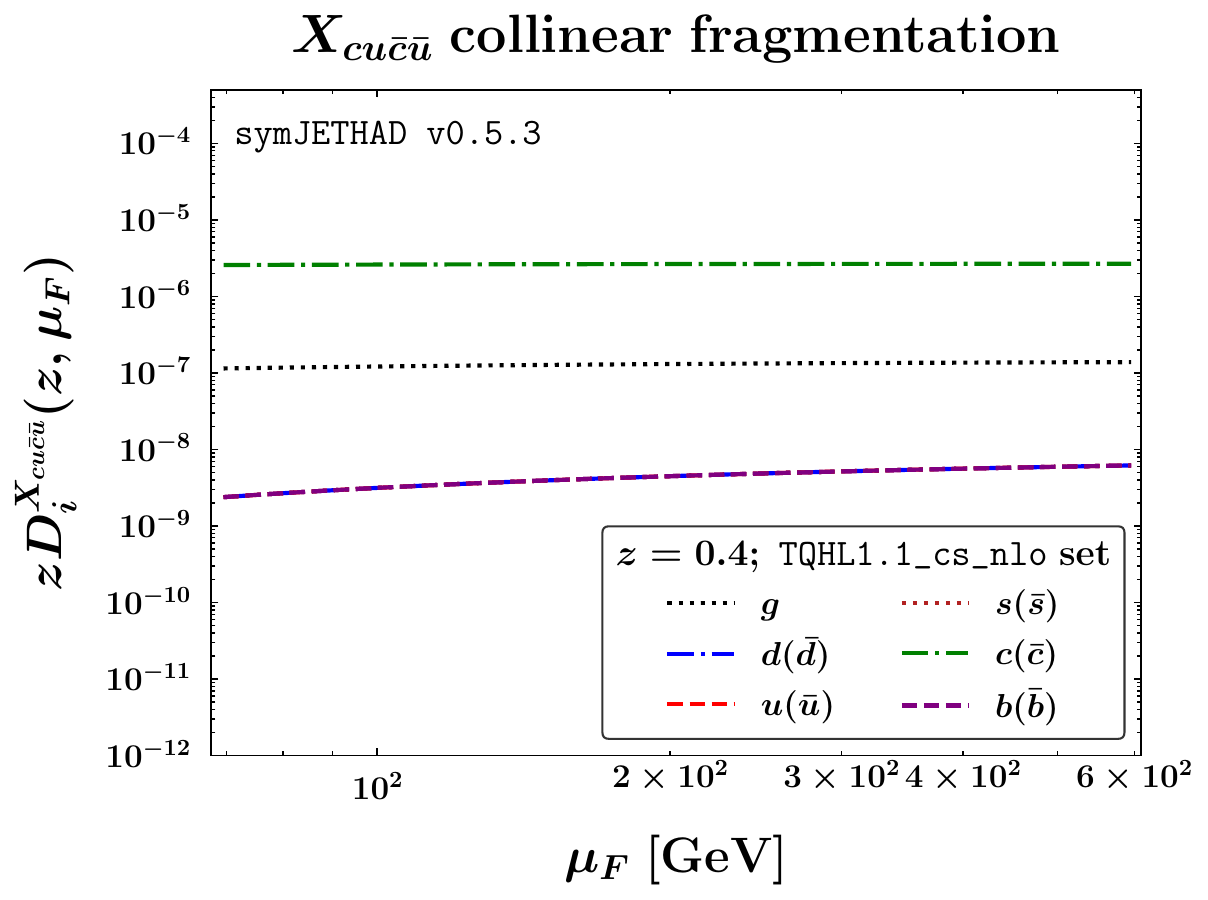}
   \hspace{-0.15cm}
   \includegraphics[scale=0.410,clip]{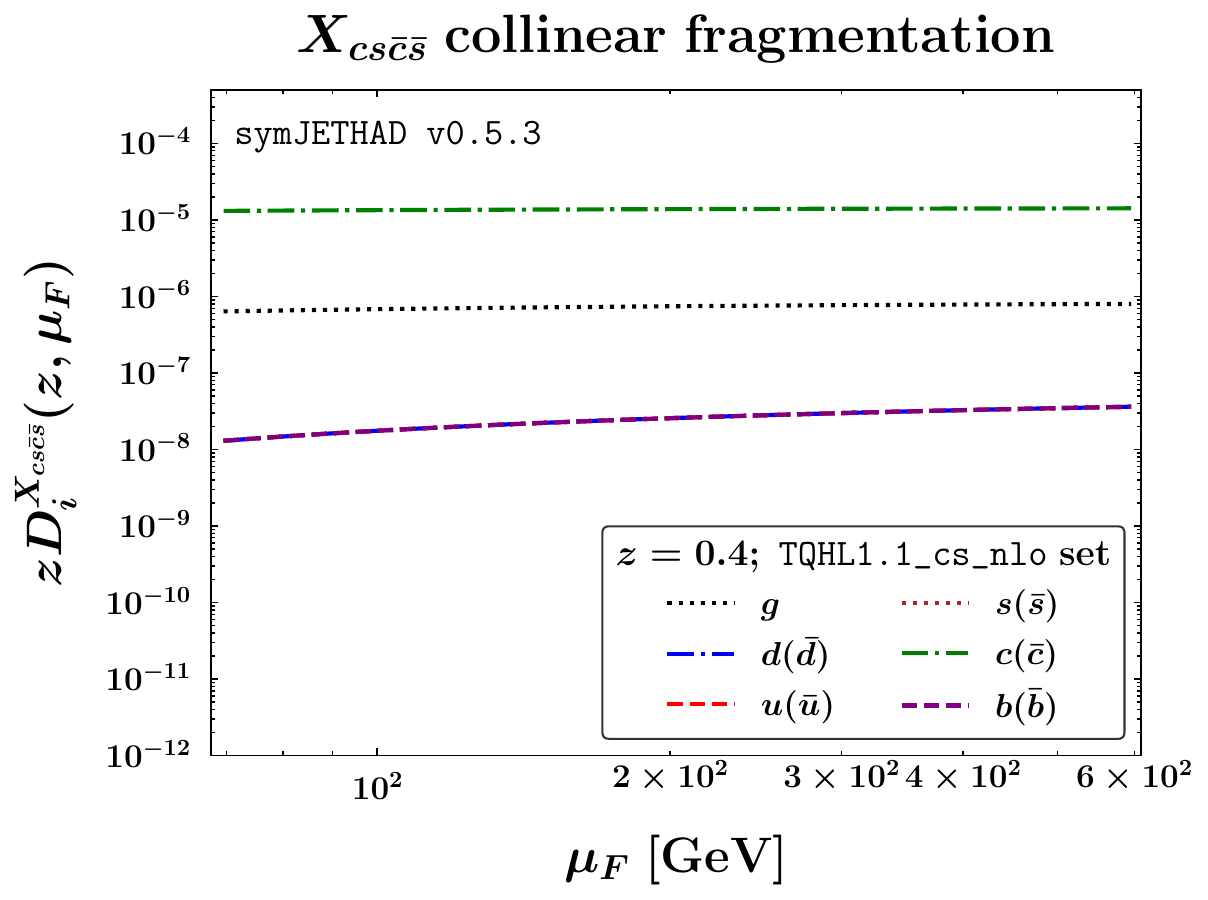}

\caption{Factorization-scale dependence of {\tt TQHL1.1} functions~\cite{Celiberto:2024beg} depicting the collinear fragmentation of $\Xcu$ (left) and $\Xcs$ (right) doubly charmed tetraquarks, at $z = 0.4 \simeq \langle z \rangle$.}
\label{fig:NLO_FFs_Xcq}
\end{figure*}

Furthermore, in alignment with previous findings for other heavy-flavored particles, the gluon to pentacharm FF exhibits a smooth, steady increase with $\mu_F$ (for the sake of comparison, in Fig.~\ref{fig:NLO_FFs_Xcq} we show analogous plots of {\tt TQHL1.1} functions~\cite{Celiberto:2024beg} for doubly heavy tetraquarks).

It recently came out that gluon FFs, exhibiting a smooth $\mu_F$-dependence, serve as \emph{stabilizers} for high-energy distributions sensitive to the semi-inclusive production of singly~\cite{Celiberto:2021dzy,Celiberto:2021fdp} or multiply~\cite{Celiberto:2022dyf,Celiberto:2022keu,Celiberto:2023rzw} heavy-flavored hadrons. This remarkable feature is referred to as the \emph{natural stability}~\cite{Celiberto:2022grc} of high-energy resummation (see Sec.~\ref{ssec:resummation}). 
The natural stability resulting from heavy-flavor fragmentation will play a special role in our phenomenology (see Sec.~\ref{sec:phenomenology}).

\section{Hybrid collinear and high-energy factorization}
\label{sec:hybrid_factorization}

In the initial part of this section (\ref{ssec:resummation}), we provide a concise overview of recent phenomenological developments in exploring the semi-hard regime of QCD.  
Subsequently, in the second part (\ref{ssec:NLL_cross_section}), we outline the construction of the $\NLLp$ hybrid factorization and its application to the study of semi-inclusive $\PQc$ plus jet hadroproduction.

\subsection{Progress in semi-hard phenomenology}
\label{ssec:resummation}

The production of hadrons containing heavy quarks provides a critical avenue for exploring high-energy QCD. In such processes, energy logarithms grow significantly, affecting the all-order running-coupling expansion and challenging the convergence of perturbative QCD. The Balitsky--Fadin--Kuraev--Lipatov (BFKL) formalism~\cite{Fadin:1975cb,Balitsky:1978ic} systematically resums these logarithms to all orders. 
This resummation applies to both LL and NLL, addressing terms proportional to $[\alpha_s \ln (s)]^n$ and $\alpha_s [\alpha_s^n \ln (s)]^n$, respectively.

BFKL-resummed cross sections for hadronic processes are expressed as a transverse-mo\-mentum high-energy's convolution of a universal Green's function, computed at NLO~\cite{Fadin:1998py,Ciafaloni:1998gs}, with two process-specific forward-production impact factors, also known as singly off-shell emission functions. 
These impact factors encapsulate collinear components such as parton distribution functions (PDFs) and fragmentation functions (FFs), embedding collinear factorization into the broader BFKL formalism, referred to as a \emph{hybrid} approach.

The BFKL resummation has been extensively tested in numerous phenomenological studies, often with $\NLLp$ precision. 
Notable examples include analyses of Mueller--Navelet jets~\cite{Ducloue:2013hia,Celiberto:2015yba,Celiberto:2016ygs,Celiberto:2022gji,Caporale:2018qnm,Baldenegro:2024ndr}, di-hadron and hadron-jet correlations~\cite{Celiberto:2016hae,Celiberto:2017ptm,Celiberto:2020rxb,Celiberto:2022rfj,Bolognino:2018oth,Celiberto:2020wpk,Celiberto:2022kxx}, forward Higgs production~\cite{Celiberto:2020tmb,Celiberto:2023rtu,Celiberto:2023uuk,Celiberto:2023eba,Celiberto:2023nym,Celiberto:2022zdg}, Drell–Yan processes~\cite{Celiberto:2018muu,Golec-Biernat:2018kem}, and heavy-hadron production~\cite{Celiberto:2021dzy,Celiberto:2021fdp,Celiberto:2022dyf,Celiberto:2023fzz,Celiberto:2022grc,Celiberto:2022keu,Celiberto:2024omj,Celiberto:2025ogy,Celiberto:2025euy}.
Furthermore, single-forward emissions offer a direct probe of the small-$x$ gluon content of the proton through the unintegrated gluon distribution (UGD), governed by the BFKL kernel. 
Phenomenological studies of UGDs have been conducted through exclusive vector-meson production at HERA~\cite{Bolognino:2018rhb} and the EIC~\cite{Bolognino:2021niq}.

Incorporating UGD information has enhanced collinear-factorization studies, supporting the derivation of small-$x$ resummed PDFs~\cite{Ball:2017otu,Abdolmaleki:2018jln,Bonvini:2019wxf} and enabling the development of improved low-$x$ TMDs~\cite{Bacchetta:2020vty,Bacchetta:2024fci}. 

High-energy observables sensitive to the production of heavy-flavored hadrons, such as $\Lambda_c$ baryons~\cite{Celiberto:2021dzy} and $b$-hadrons~\cite{Celiberto:2021fdp}, have demonstrated the potential to resolve challenges in describing semi-hard final states at natural scales. 
Unlike light-particle production, which suffers from large NLL corrections and nonresummed threshold logarithms~\cite{Bolognino:2018oth,Celiberto:2020wpk}, heavy-hadron emissions exhibit a \emph{natural stabilization} behavior~\cite{Celiberto:2022grc}. 
This stabilization reflects the dominant role of collinear fragmentation in describing high-$p_T$ heavy-hadron dynamics.

Building on this natural stability, novel DGLAP-evolving FFs have been developed using NRQCD inputs~\cite{Braaten:1993mp,Zheng:2019dfk,Braaten:1993rw,Zheng:2019gnb}, initially for vector quarkonia~\cite{Celiberto:2022dyf,Celiberto:2023fzz} and later for $B_c^{(*)}$ states~\cite{Celiberto:2022keu,Celiberto:2024omj}. The stability observed in these studies has also fostered connections with exotic hadron physics~\cite{Celiberto:2023rzw,Celiberto:2024mrq,Celiberto:2024mab,Celiberto:2024beg}.

\begin{figure*}[!t]
\centering
\includegraphics[width=0.75\textwidth]{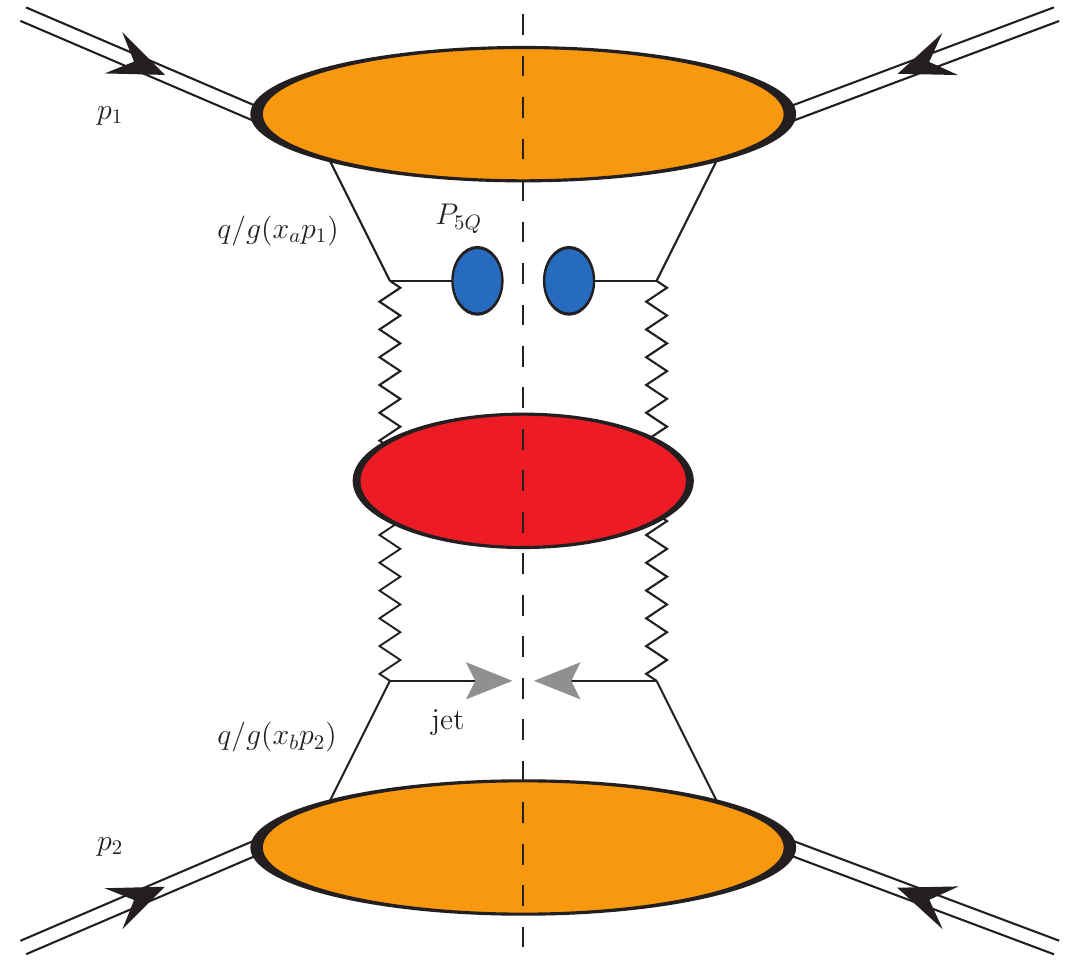}

\caption{Hybrid collinear and high-energy factorization for the $\PQQ$ plus jet semi-inclusive detection at hadron colliders.
Blue ovals describe heavy-pentaquark
collinear FFs. Gray arrows represent light-flavored jets. 
Orange blob depict proton collinear PDFs.
The BFKL Green’s function (red oval) is connected to the two impact factors by Reggeon lines. Diagrams made with {\tt JaxoDraw 2.0}~\cite{Binosi:2008ig}.}
\label{fig:reaction}
\end{figure*}

\subsection{NLL/NLO$^+$ cross section}
\label{ssec:NLL_cross_section}

We study the following hadroproduction (see also Fig.~\ref{fig:reaction})
\begin{equation}
\label{process}
    {\rm p}(p_1) + {\rm p}(p_2) \, \rightarrow \, \PQc(\kappa_1, y_1, \varphi_1) + {\cal X} + {\rm jet}(\kappa_2, y_2, \varphi_2) \; ,
\end{equation}
where ${\rm p}(p_{1,2})$ is an incoming proton with momentum $p_{1,2}$.
Then, $\PQc$ is $| c \bar{c} c c c \rangle$ pentacharm emitted with momentum $\kappa_1$, rapidity $y_1$, and azimuthal angle $\varphi_1$.
A light-flavored jet is simultaneously detected with momentum $\kappa_2$, rapidity $y_2$, and azimuthal angle $\varphi_2$.
Finally, ${\cal X}$ inclusively depict all the undetected products. 
Large transverse momenta, $|\vec \kappa_{1,2}|$, and large rapidity separations, $\DY = y_1 - y_2$, make our final states semi-hard and diffractive.
Moreover, large transverse momenta are needed to ensure that the ZM-VFNS treatment of the leading-power fragmentation is the dominant formation mechanism of our $\PQc$ states.

The momenta of the parent protons constitute a Sudakov basis with $p_1^2= P_b^2=0$ and $(p_1\cdot P_b) = s/2$, so that
\begin{equation}\label{sudakov}
\kappa_{1,2} = x_{1,2} p_{1,2} - \frac{ \kappa_{1,2\perp}^2}{x_{1,2} s}p_{2,1} + \kappa_{1,2\perp} \ , \quad
\vec \kappa_{1,2}^{\,2} \equiv -\kappa_{1,2\perp}^2\;.
\end{equation}
The relations
\begin{equation}\label{y-vs-x}
y_{1,2}=\pm\frac{1}{2}\ln\frac{x_{1,2}^2 s}
{\vec \kappa_{1,2}^2 }
\qquad \mbox{and} \qquad
\drv y_{1,2} = \pm \frac{\drv x_{1,2}}{x_{1,2}}
\end{equation}
hold between longitudinal fractions ($x_{1,2}$) and rapidities ($y_{1,2}$) of the observed particles.
Then we have
\begin{equation}
\label{DeltaY}
 \DY = y_1 - y_2 = \ln \left( \frac{x_1 x_2}{|\vec \kappa_1||\vec \kappa_2|} s \right) \;.
\end{equation}

In a pure LO QCD collinear factorization framework, the differential cross section for our process is expressed as a collinear convolution of the on-shell hard subprocess term, the proton PDFs, and the pentacharm FFs
\begin{equation}
\label{sigma_collinear}
\begin{split}
\hspace{-0.25cm}
\frac{\drv\sigma^{\rm LO}_{\rm [coll.]}}{\drv x_1\drv x_2\drv ^2\vec \kappa_1\drv ^2\vec \kappa_2}
= \hspace{-0.25cm} \sum_{\alpha,\beta=q,{\bar q},g}\int_0^1 \hspace{-0.20cm} \drv x_a \int_0^1 \hspace{-0.20cm} \drv x_b\ f_\alpha\left(x_a\right) f_\beta\left(x_b\right)
\int_{x_1}^1 \hspace{-0.15cm} \frac{\drv \xi}{\xi}D^{\PQc}_\alpha\left(\frac{x_1}{\xi}\right) 
\frac{\drv {\hat\sigma}_{\alpha,\beta}\left(\hat s\right)}
{\drv x_1\drv x_2\drv ^2\vec \kappa_1\drv ^2\vec \kappa_2}\;,
\end{split}
\end{equation}
where the indices $\alpha, \beta$ sum over all parton species except for the top quark, which does not participate in fragmentation.\footnote{For conciseness, the explicit dependence on $\mu_F$ has been omitted from Eq.~\eqref{sigma_collinear}.}  
In this expression, $f_{\alpha,\beta}\left(x_{a,b}, \mu_F \right)$ represent the PDFs of the proton, while $D^{\PQc}_\alpha\left(x_1/\xi, \mu_F \right)$ describe the fragmentation of a parton into the pentaquark.  
The quantities $x_{a,b}$ refer to the longitudinal momentum fractions carried by the incoming partons, and $\xi$ indicates the momentum fraction associated with the outgoing parton that generates the exotic hadron.  
Finally, $\drv\hat\sigma_{\alpha,\beta}\left(\hat s \right)$ denotes the partonic collinear hard factors, where $\hat s \equiv x_a x_b s$ represents the center-of-mass energy squared for the partonic subprocess.

On the other hand, the high-energy resummed differential cross section within the hybrid factorization framework is represented as a transverse-momentum convolution involving the BFKL Green's function and two impact factors.   
The differential cross section can further be decomposed into a Fourier series expansion with respect to the azimuthal angle difference, $\varphi = \varphi_1 - \varphi_2 - \pi$, to write
\begin{equation}
 \label{dsigma_Fourier}
 \frac{\drv \sigma}{\drv y_1 \drv y_2 \drv \vec \kappa_1 \drv \vec \kappa_2 \drv \varphi_1 \drv \varphi_2} =
 \frac{1}{(2\pi)^2} \left[{\cal C}_0 + 2 \sum_{m=1}^\infty \cos (m \varphi)\,
 {\cal C}_m \right]\, .
\end{equation}
The first key ingredient is NLL Green's function
\begin{equation}
\label{G_BFKL_NLL}
 {\cal G}_{\rm NLL}(\DY,m,\nu,\mu_R) = e^{{\DY} \bar \alpha_s(\mu_R) \,
 \chi^{\rm NLO}(m,\nu)} \; ,
\end{equation}
with $\bar \alpha_s(\mu_R) \equiv \alpha_s(\mu_R) N_c/\pi$ and $\beta_0 = 11N_c/3 - 2 n_f/3$ being the leading coefficient of the QCD $\beta$-function.
The $\chi$ quantity entering the exponent of Eq.~\eqref{G_BFKL_NLL} is the BFKL kernel in the Mellin space, which embodies the resummation of energy logarithms at NLL.
It reads
\begin{eqnarray}
 \label{chi}
 \chi^{\rm NLO}(m,\nu) = \chi(m,\nu) + \bar\alpha_s \tilde{\chi}(m,\nu) \;,
\end{eqnarray}
where $\chi(l,\nu)$ are the LO kernel eigenvalues
\begin{eqnarray}
 \label{kernel_LO}
 \chi\left(m,\nu\right) = -2\gamma_{\rm E} - 2 \, {\rm Re} \left\{ \psi\left(\frac{1}{2} + \frac{m}{2} + i \nu \right) \right\} \, ,
\end{eqnarray}
$\gamma_{\rm E}$ is the Euler--Mascheroni constant and $\psi(z) \equiv \Gamma^\prime
(z)/\Gamma(z)$ stands for the logarithmic derivative of the Gamma function. 
The $\tilde{\chi}(m,\nu)$ function in Eq.\eref{chi} is the NLO correction to the kernel
\begin{equation}
\label{chi_NLO}
\tilde{\chi} \left(m,\nu\right) = \bar\chi(m,\nu)+\frac{\beta_0}{2 N_c}\chi(m,\nu)
\left( - \frac{1}{4}\chi(l,\nu) + \frac{5}{6} + \ln\frac{\mu_R}{\sqrt{|\vec \kappa_1||\vec \kappa_2|}} \right) \;,
\end{equation}
with $\bar\chi(m,\nu)$ function being computed in Ref.~\cite{Kotikov:2000pm}.

Another fundamental element for building our resummed differential cross section is the pentacharm NLO emission function, projected onto the LO kernel eigenfunctions. 
This emission function is based on the formulation presented in Ref.~\cite{Ivanov:2012iv}, which is particularly suited for exploring the dynamics of heavy hadrons produced at large transverse momentum
\begin{equation}
\label{PIF}
\E_{\PQc}^{\rm NLO}(m,\nu,|\vec \kappa|,x) =
\E_{\PQc}(\nu,|\vec \kappa|,x) +
\alpha_s(\mu_R) \, \hat \E_{\PQc}(m,\nu,|\vec \kappa|,x) \;.
\end{equation}
Its LO expression reads
\begin{equation}
\label{LOPIF}
\hspace{-0.30cm}
\E_{\PQc}(\nu,|\vec \kappa|,x) 
= 2 \sqrt{\frac{C_F}{C_A}}
|\vec \kappa|^{2i\nu-1}
\!\!\!\int_{x}^1\frac{\drv \xi}{\xi}
\left( \frac{\xi}{x} \right)
^{2 i\nu-1} 
 \!\left[\frac{C_A}{C_F}f_g(\xi)D_g^{\PQc}\left(\frac{x}{\xi}\right)
 +\!\!\!\sum_{\alpha=q,\bar q}\!f_\alpha(\xi)D_\alpha^{\PQc}\left(\frac{x}{\xi}\right)\right] \;.
\end{equation}
Its NLO correction, $\hat \E_{\PQc}(m,\nu,|\vec \kappa|,x)$, can be found in Ref.~\cite{Ivanov:2012iv}.

The last ingredient is the light-flavored jet impact factor
\begin{equation}
\label{JIF}
\E_J^{\rm NLO}(m,\nu,|\vec \kappa|,x) =
\E_J(\nu,|\vec \kappa|,x) +
\alpha_s(\mu_R) \, \hat \E_J(m,\nu,|\vec \kappa|,x) \;,
\end{equation}
whose NLO limit reads
\begin{equation}
 \label{LOJIF}
 \E_J(\nu,|\vec \kappa|,x) = 2 \sqrt{\frac{C_F}{C_A}}
 |\vec \kappa|^{2i\nu-1}\left[\frac{C_A}{C_F}f_g(x)
 +\sum_{\beta=q,\bar q}f_\beta(x)\right] \;.
\end{equation}
Its NLO correction, $\hat \E_J(m,\nu,|\vec \kappa|,x)$, depends on the jet algorithm. 
We follow a suitable strategy, given by combining Eq.~(36) of Ref.~\cite{Caporale:2012ih} with Eqs.~(4.19) and~(4.20) of Ref.~\cite{Colferai:2015zfa}.
This approach builds upon calculations from Refs.~\cite{Ivanov:2012ms}, optimized for numerical analyses. 
In these works, a jet selection function was derived using the small-cone approximation (SCA) adapted to the cone-type algorithm (we refer the reader to Ref.~\cite{Colferai:2015zfa} for technical details). 
Consistent with the choice made in recent CMS experimental studies on forward-jet events~\cite{Khachatryan:2016udy}, we adopt a jet-cone radius of $r_J = 0.5$.

We bring together all the necessary components to formulate our master equation for the $\NLLp$ azimuthal coefficients, expressed in the $\MSb$ renormalization scheme. 
For further technical details, we refer to Ref.~\cite{Caporale:2012ih}.
We write

\begin{eqnarray}
\label{Cm_NLLp_MSb}
 \CmNLLp \!\! &=& \!\! 
 \frac{e^{\DY}}{s} 
 \int_{-\infty}^{+\infty} \drv \nu \, 
 {\cal G}_{\rm NLL}(\DY,m,\nu,\mu_R) \,
 \alpha_s^2(\mu_R) 
 \\ \nonumber
 \!\! &\times& \!\! \biggl\{\E_{\PQc}^{\rm NLO}(m,\nu,|\vec \kappa_1|, x_1) \,
 [\E_J^{\rm NLO}(m,\nu,|\vec \kappa_2|,x_2)]^*
 \\ \nonumber
 \!\! &+& \!\!
 \left.
 \alpha_s^2(\mu_R) \DY \frac{\beta_0}{4 \pi} \,
 \chi(m,\nu)
 \left[\ln\left(|\vec \kappa_1| |\vec \kappa_2|\right) + \frac{i}{2} \, \frac{\drv}{\drv \nu} \ln\frac{\E_{\PQc}}{\E_J^*}\right]
 \right\}
 \;.
\end{eqnarray}
The label $\NLLp$ means a full NLL resummation of energy logarithms while maintaining NLO perturbative accuracy. 
The `$+$' superscript highlights the inclusion of contributions beyond the NLL level, which stem from cross terms involving NLO corrections to the emission functions in our representation of the azimuthal coefficients.

In our framework, the notation $\NLLpp$ reflects the perturbative accuracy of the universal BFKL kernel employed in the resummation of high-energy logarithms. Specifically, the $\LL$ level corresponds to resumming terms of the form $[\alpha_s \ln(s)]^n$ via the LO BFKL kernel, while the $\NLLpp$ level includes contributions of type $\alpha_s [\alpha_s \ln(s)]^n$, captured by the NLO kernel. 
The kernel is then convoluted with process-dependent impact factors, computed at LO or NLO to match the desired fixed-order accuracy of the observable. 
This kernel-based counting provides a universal prescription for counting logarithmic accuracy across different observables.
By referring to the accuracy of the BFKL kernel, rather than to the overall power counting in the cross section, one avoids ambiguities that would otherwise arise due to the process-dependent nature of the impact factors.
As an example, the leading-$\alpha_s$ power of the impact factor differs between Higgs production and jet or hadron production~\cite{Celiberto:2020tmb}.

For comparison purposes, we will also examine the pure LL limit within the $\MSb$ scheme.
\begin{equation}
\label{Cm_LL_MSb}
 \CmLL = 
 \frac{e^{\DY}}{s} 
 \int_{-\infty}^{+\infty} \drv \nu \, 
 e^{{\cal G}_{\rm NLL}^{(0)}(\DY,m,\nu,\mu_R)} 
 \alpha_s^2(\mu_R) \, \E_{\PQc}(m,\nu,|\vec \kappa_1|, x_1)[\E_J(m,\nu,|\vec \kappa_2|,x_2)]^* \;.
\end{equation}

To effectively compare high-energy resummed and DGLAP predictions, it is necessary to evaluate observables using both our hybrid factorization and pure fixed-order computations.
However, as far as we know, no numerical code currently exists for fixed-order distributions in inclusive semi-hard hadron-plus-jet hadroproduction processes at NLO accuracy. 
To address this limitation and evaluate the impact of high-energy resummation on DGLAP predictions, we will compare our BFKL-based results with those derived using a high-energy fixed-order approach.

This method, originally developed for light di-jet~\cite{Celiberto:2015yba} and hadron-jet~\cite{Celiberto:2020wpk} azimuthal correlations, truncates the high-energy series at NLO accuracy, effectively reproducing the high-energy behavior of a pure NLO calculation.
In practical terms, we limit the expansion of the azimuthal coefficients in Eq.~(\ref{Cm_NLLp_MSb}) to ${\cal O}(\alpha_s^3)$, yielding an effective high-energy fixed-order ($\HENLOp$) expression suited for our phenomenological study.

The $\MSb$ formula for our angular coefficients within the $\HENLOp$ expansion reads
\begin{align}
\label{Cm_HENLOp_MSb}
 \ClHENLOp &= 
 \frac{e^{\DY}}{s} 
 \int_{-\infty}^{+\infty} \drv \nu \, 
 \alpha_s^2(\mu_R) \,
 \left[ 1 + {\cal G}_{\rm NLL}^{(0)}(\DY,m,\nu,\mu_R) \right]
 \\ \nonumber
 &\times
 \E_{\PQc}^{\rm NLO}(m,\nu,|\vec \kappa_1|, x_1)[\E_J^{\rm NLO}(m,\nu,|\vec \kappa_2|,x_2)]^* \;,
\end{align}
with
\begin{equation}
\label{G_BFKL_0}
 {\cal G}_{\rm NLL}^{(0)}(\DY,l,\nu,\mu_R) = \bar \alpha_s(\mu_R) \DY \chi(l,\nu)
\end{equation}
being the expansion of the BFKL kernel at the first order in $\alpha_s$.

We base our choice of values of the factorization scale ($\mu_F$) and the renormalization scale ($\mu_R$) upon the \emph{natural} energy scales provided by the final state. 
Specifically, we set $\mu_F = \mu_R = \mu_N$, where $\mu_N = M_{\PQc \perp} + |\vec \kappa_2|$ serves as the process-natural reference scale. Here, $M_{\PQc \perp} = \sqrt{M_{\PQc}^2 + |\vec \kappa_1|^2}$ represents the transverse mass of the pentacharm, with $M_{\PQc}$ its mass.

In principle, any scale comparable in magnitude to the typical energy of the emitted particle can be regarded as a \emph{natural} choice. For our process, the emission of two particles suggests two natural scales: $M_{\PQc \perp}$ for the pentacharm and $|\vec \kappa_2|$ for the jet. 
To facilitate comparisons with other approaches, we adopt a simplified scheme that combines these into a single scale, specifically the sum of $m_{1 \perp}$ and $m_{2 \perp}$. This choice aligns with the conventions used in several numerical codes designed for precision QCD studies (see, \emph{e.g.}, Refs.~\cite{Alioli:2010xd,Campbell:2012am}).

To estimate the magnitude of missing higher-order uncertainties (MHOUs), the $\mu_F$ and $\mu_R$ scales will be varied uniformly within the range $\mu_N/2$ to $2\mu_N$, controlled by the $C_\mu$ parameter (see Sec.~\ref{sec:phenomenology}).

\section{Phenomenology at next-generation hadron colliders}
\label{sec:phenomenology}

All numerical results presented in this work were obtained using the \textsc{Python}+\textsc{Fortran} {\Jethad} multimodular interface~\cite{Celiberto:2020wpk,Celiberto:2022rfj,Celiberto:2023fzz,Celiberto:2024mrq,Celiberto:2024swu}. 
For proton PDFs, we utilized the {\tt NNPDF4.0} NLO set~\cite{NNPDF:2021uiq}, accessed through {\tt LHAPDF v6.5.5}~\cite{Buckley:2014ana}. 
The uncertainty bands displayed in the plots account for the combined effects of MHOUs and errors from multidimensional numerical integrations, which were consistently maintained below 1\% by the {\Jethad} integrators.

\subsection{Observables and final states}
\label{ssec:obserbables}

We will examine two primary observables: the rapidity-interval distribution, which represents the cross section differential with respect to the rapidity separation, $\Delta Y = y_1 - y_2$, between the two produced particles, and the $\vec{\kappa}_1$-differential transverse-momentum distribution.

The first observable is directly connected to the ${\cal C}_0$ angular coefficient, as defined in Sec.~\ref{ssec:NLL_cross_section}, integrated over the final-state transverse momenta and rapidities, and evaluated at fixed values of the rapidity separation $\Delta Y$ between the pentaquark ant the light jet.
We write
\\
\begin{equation}
 \label{obs:I}
 \frac{\drv \sigma(\DY, s)}{\drv \DY} =
 \int_{|\vec \kappa_2|^{\rm min}}^{|\vec \kappa_2|^{\rm max}} 
 \!\!\drv |\vec \kappa_1|
 \int_{|\vec \kappa_2|^{\rm min}}^{|\vec \kappa_2|^{\rm max}} 
 \!\!\drv |\vec \kappa_2|
 \int_{\max \, (y_1^{\rm min}, \, \DY + y_2^{\rm min})}^{\min \, (y_1^{\rm max}, \, \DY + y_2^{\rm max})} \drv y_1
 \, \,
 {\cal C}_0^{\rm [res]}
\Bigm \lvert_{y_2 \;=\; y_1 - \DY}
 \;,
\end{equation}
\\
where the `${\rm [res]}$' label for ${\cal C}_0$ inclusively denotes $\NLLp$, $\HENLOp$, or $\LL$.  
To remove the integration over one of the two rapidities, say $y_2$, we used a $\delta(\Delta Y - (y_1 - y_2))$ function.  

The transverse momenta of the forward $\PQc$ state are restricted within the $60 < |\vec \kappa_1|/{\rm GeV} < 120$ range. 
These cuts align well with the conditions required for the applicability of a ZM-VFNS fragmentation approach, where energy scales must be sufficiently higher than the thresholds for the DGLAP evolution of heavy-quark species.  
The jet is tagged in transverse-momentum ranges slightly different but compatible with current LHC and upcoming HL-LHC studies~\cite{Khachatryan:2016udy}, such as $50 < |\vec \kappa_2|/{\rm GeV} < 60$. 

The use of such \emph{asymmetric} transverse-momentum ranges facilitates disentangling pure high-energy dynamics from fixed-order signals~\cite{Celiberto:2015yba,Celiberto:2020wpk}. 
Moreover, it mitigates the large Sudakov logarithms associated with nearly back-to-back configurations, which would otherwise require an additional resummation~\cite{Mueller:2013wwa,Marzani:2015oyb}.  
Finally, this selection suppresses instabilities related to radiative corrections~\cite{Andersen:2001kta} and avoids violations of energy-momentum  conservation~\cite{Ducloue:2014koa}.

Regarding rapidity ranges, we adopt proxy cuts from current LHC studies. 
Pentacharms are detected exclusively within the barrel calorimeter region~\cite{Chatrchyan:2012xg}, spanning from $-2.5$ to $2.5$, while light-flavored jets can also be tagged in the endcap regions~\cite{Khachatryan:2016udy}, extending from $-4.7$ to $4.7$.

The second observable is the following transverse-momentum rate
\begin{equation}
\label{obs:I-k1b}
 \frac{\drv \sigma(|\vec \kappa_1|, s)}{\drv |\vec \kappa_1|} =
 \int_{|\vec \kappa_2|^{\rm min}}^{|\vec \kappa_2|^{\rm max}} 
 \!\!\drv |\vec \kappa_2|
 \, \,
 \int_{\DY^{\rm min}}^{\DY^{\rm max}} \drv \DY
 \int_{\max \, (y_1^{\rm min}, \, \DY + y_2^{\rm min})}^{\min \, (y_1^{\rm max}, \, \DY + y_2^{\rm max})} \drv y_1
 \, \,
 {\cal C}_{l=0}^{\rm [res]}
\Bigm \lvert_{y_2 \;=\; y_1 - \DY}
 \;.
\end{equation}
This distribution is differential in the $\PQc$ transverse momentum $|\vec \kappa_1|$, but integrated over the jet transverse-momentum range of $40~\text{GeV} < |\vec \kappa_2| < 120~\text{GeV}$ and within a specific $\DY$-bin.
Rapidity cuts for pentacharm and jet detections are the same as before.

This $|\vec \kappa_1|$-distribution provides a foundation for exploring potential connections between the $\NLLp$ factorization and alternative formalisms. 
Allowing $|\vec \kappa_1|$ to vary from $10$ to $100~\text{GeV}$ facilitates the examination of an extensive kinematic region, where additional resummation mechanisms could become significant. 
Recent investigations of high-energy Higgs~\cite{Celiberto:2020tmb} and heavy-jet~\cite{Bolognino:2021mrc} tagging have shown that our hybrid approach is reliable within the moderate transverse-momentum regime, specifically when $|\vec \kappa_1|$ and $|\vec \kappa_2|$ are of comparable magnitude. 

On the other hand, large $|\vec \kappa_1|$ values are expected to enhance threshold logarithms in the hard regime ($|\vec \kappa_1| > |\vec \kappa_2|^{\rm max}$), while soft logarithms become prominent in the low-momentum regime ($|\vec \kappa_1| \ll |\vec \kappa_2|^{\rm min}$). 
Consequently, studying $ |\vec \kappa_1| $-distributions serves not only as a validity check for our framework, but also as an effective way to prepare the ground for embodying additional resummation techniques.

\subsection{Numerical analysis}
\label{ssec:results}

\begin{figure*}[!t]
\centering

   \hspace{0.00cm}
   \includegraphics[scale=0.395,clip]{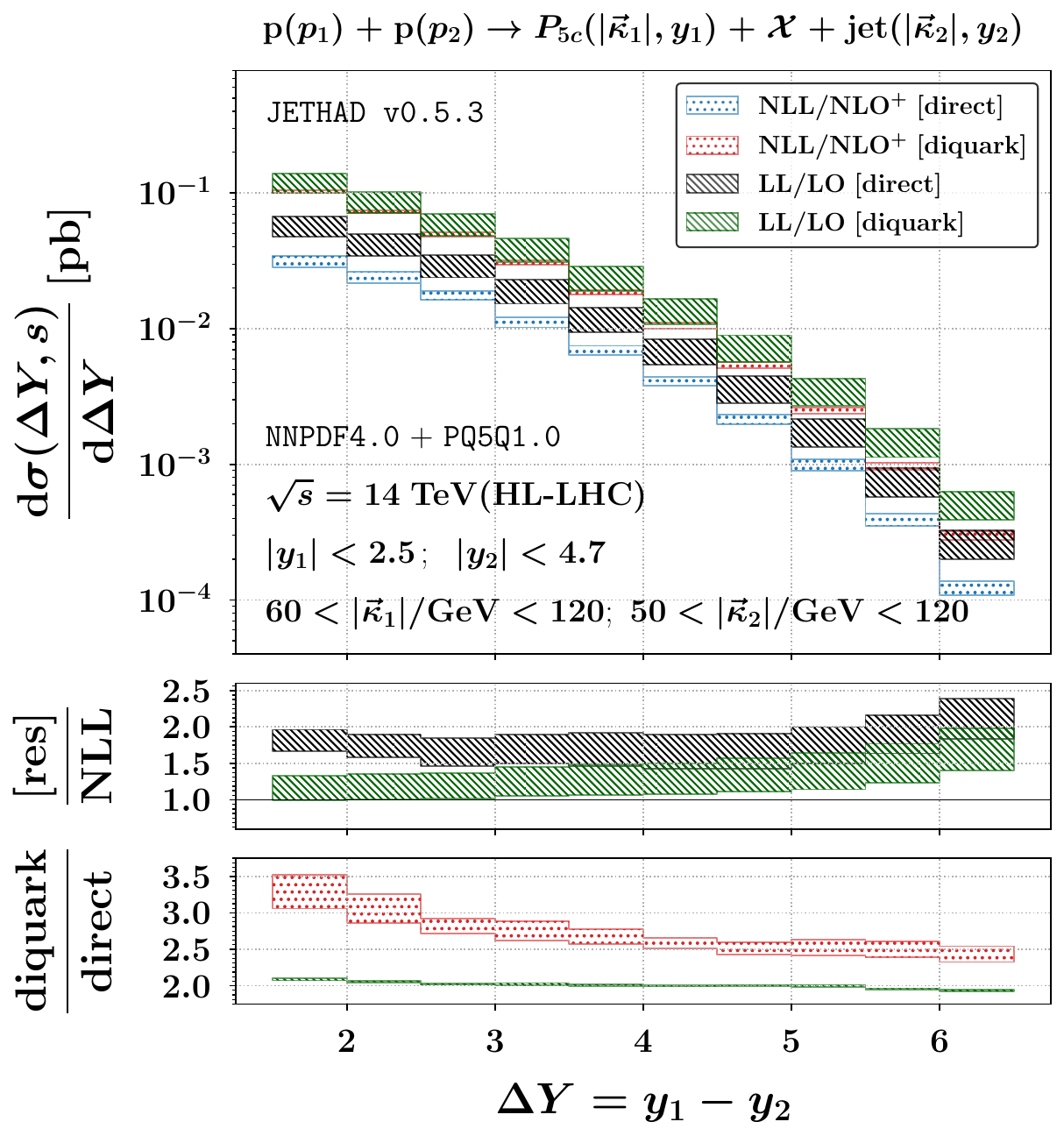}
   \hspace{0.30cm}
   \includegraphics[scale=0.395,clip]{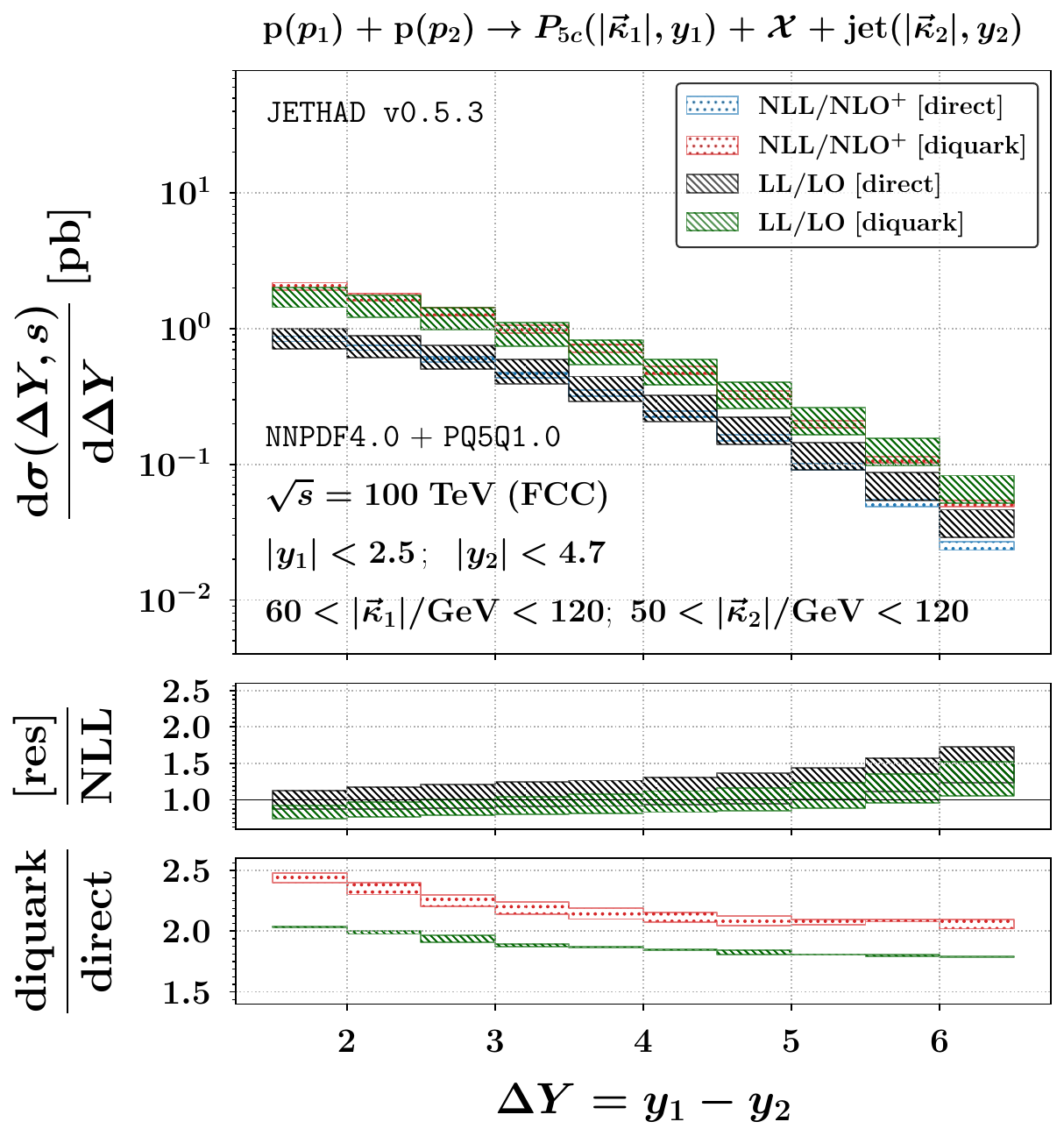}

\caption{Rapidity distributions for the semi-inclusive $\PQc$ plus light-jet production at $\sqrt{s} = 14$ TeV (HL-LHC, left) and $100$ TeV (nominal FCC, right).
First ancillary panels below primary plots show the ratio between $\LL$ and $\NLLp$.
Second ancillary panels exhibit the ratio between diquark and direct initial-scale fragmentation.
Uncertainty bands capture the net effect of MHOUs and phase-space multidimensional integration.}
\label{fig:I}
\end{figure*}

We show in Fig.~\ref{fig:I} resummed predictions for $\PQc$ plus jet $\DY$-distributions at $14$~TeV HL-LHC (left) and $100$~TeV FCC (right). 
To facilitate realistic comparisons with future experimental data, we use uniformly sized $\DY$-bins with a fixed length of $0.5$. 

The first ancillary panels below the primary plots display the ratio between pure $\LL$ results and $\NLLp$ ones.
The second ancillary panels capture the ratio between diquark and direct fragmentation channels.

The cross-section statistics are quite promising, approximately ranging from $10^{-4}$~pb to $20$~pb. 
In particular, the rates increase by roughly an order of magnitude as $\sqrt{s}$ transitions from typical HL-LHC energies to nominal FCC ones (we note that the scale of the vertical axis differs between the two plots).

\begin{figure*}[!t]
\centering

   \hspace{0.00cm}
   \includegraphics[scale=0.395,clip]{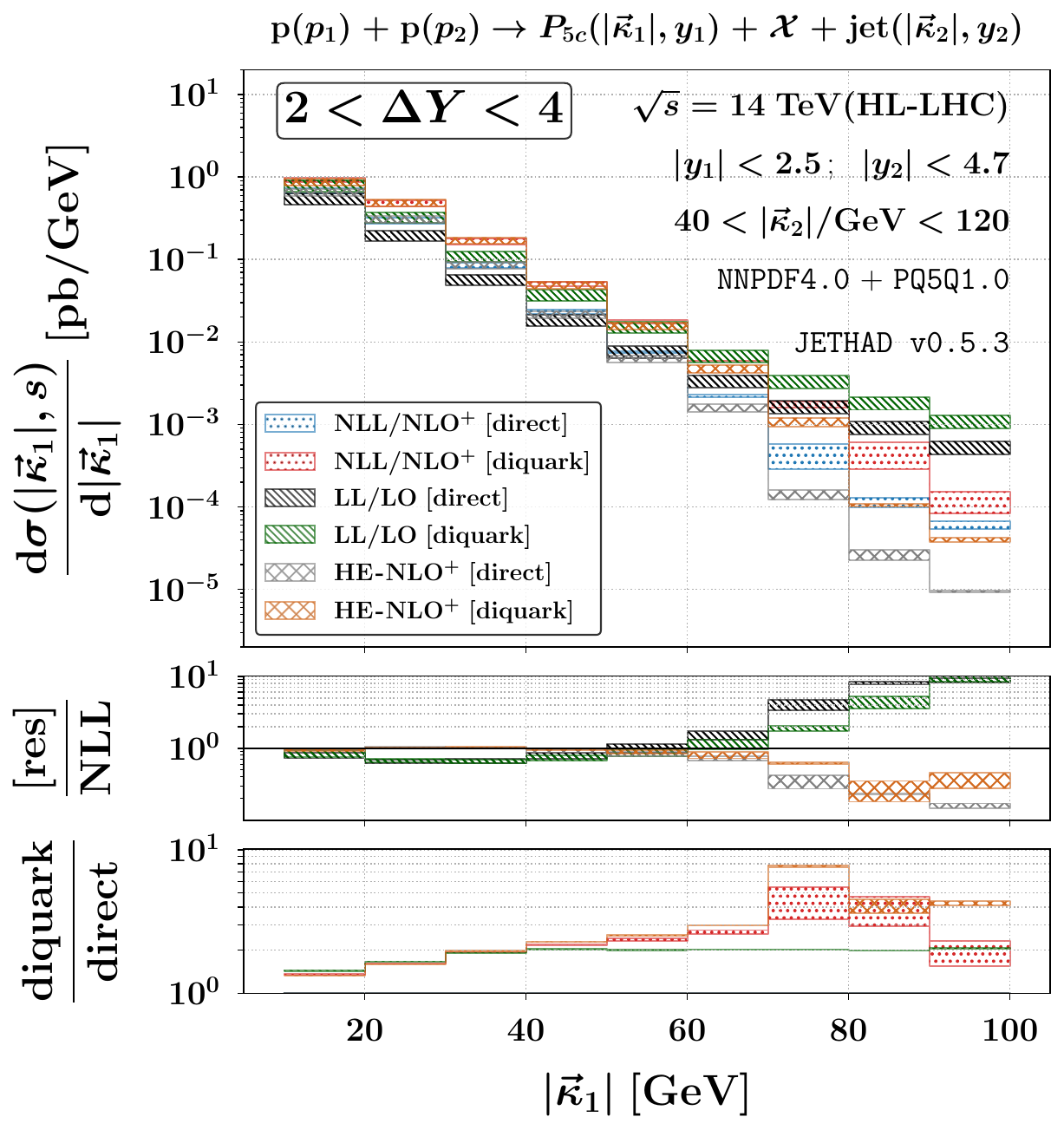}
   \hspace{0.00cm}
   \includegraphics[scale=0.395,clip]{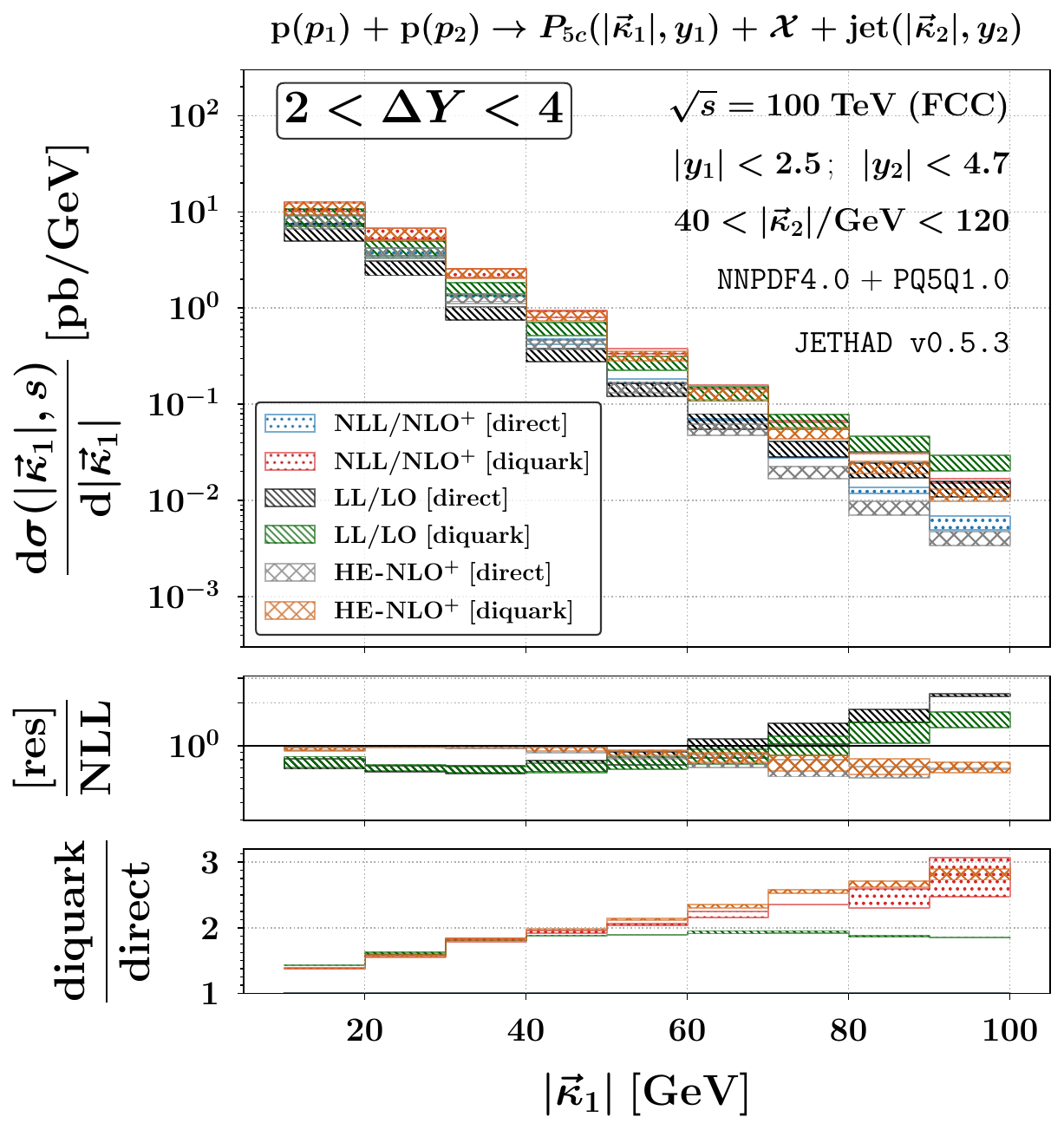}

\caption{Transverse-momentum distributions for the semi-inclusive $\PQc$ plus light-jet production at $\sqrt{s} = 14$ TeV (HL-LHC, left) and $100$ TeV (nominal FCC, right), and for $2 < \DY < 4$.
First ancillary panels below primary plots show the ratio between $\LL$ or $\HENLOp$ and $\NLLp$.
Second ancillary panels exhibit the ratio between diquark and direct initial-scale fragmentation.
Uncertainty bands capture the net effect of MHOUs and phase-space multidimensional integration.}
\label{fig:I-k1b_YS}
\end{figure*}

\begin{figure*}[!t]
\centering

   \hspace{0.00cm}
   \includegraphics[scale=0.395,clip]{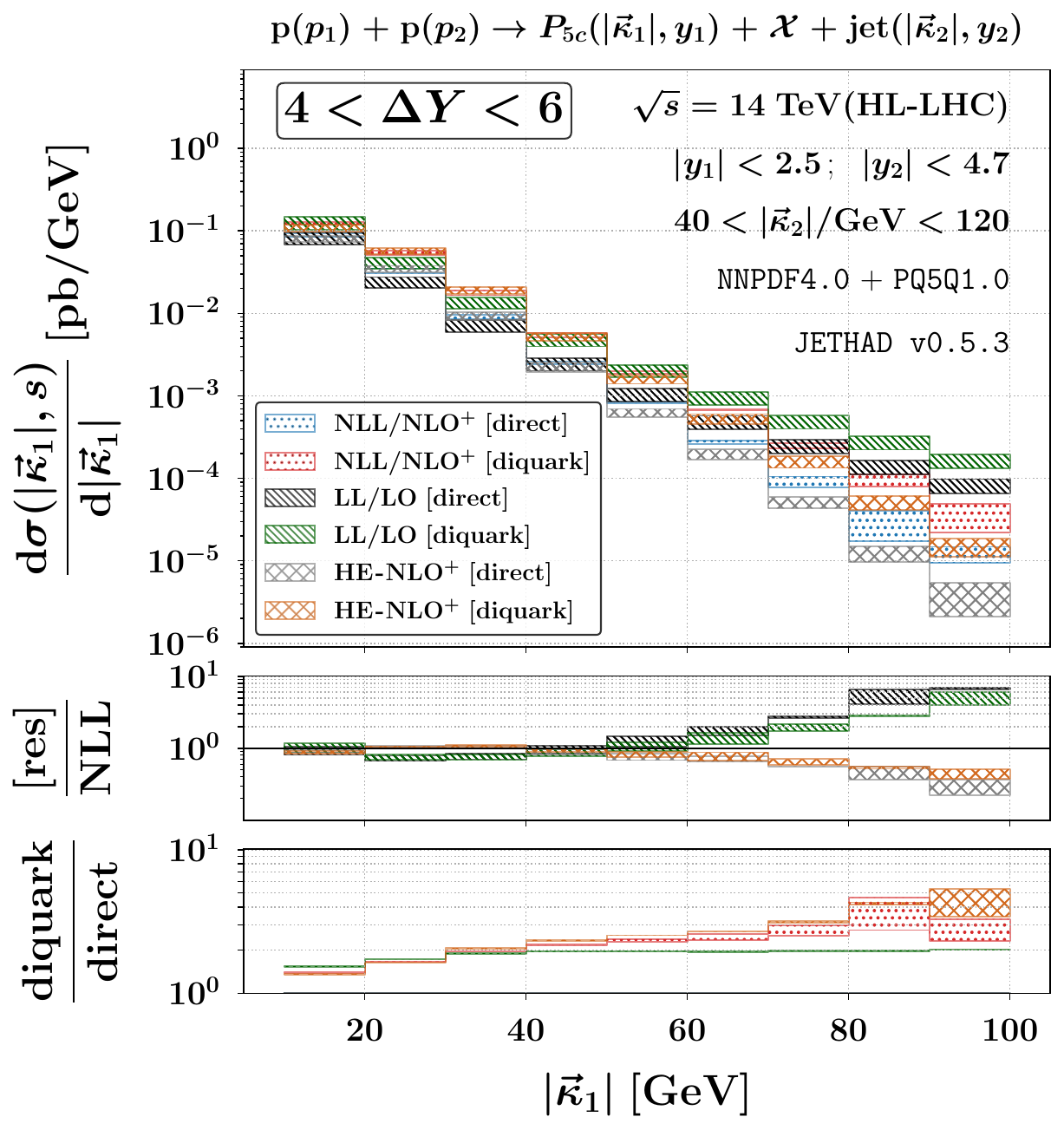}
   \hspace{0.00cm}
   \includegraphics[scale=0.395,clip]{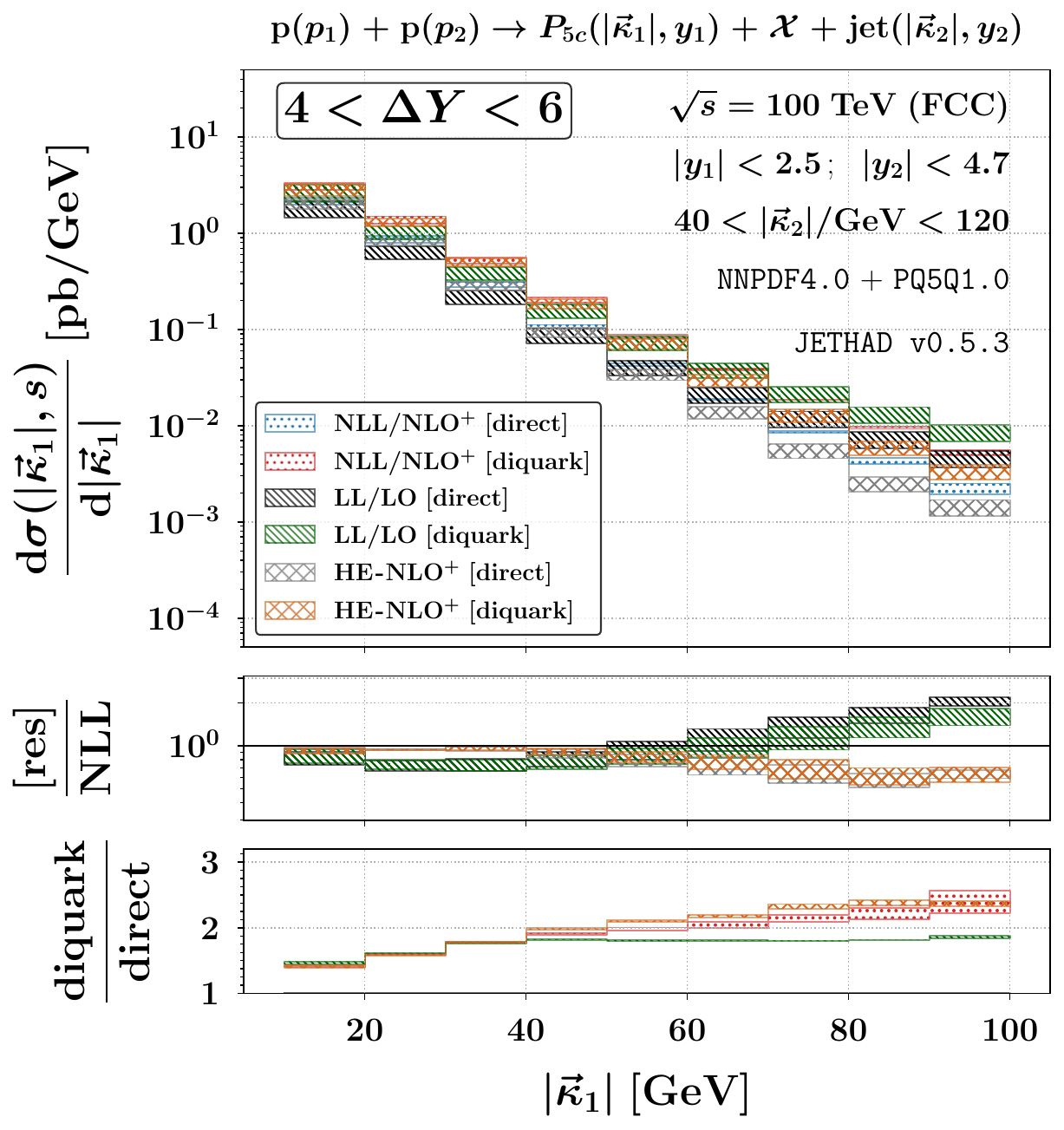}

\caption{Transverse-momentum distributions for the semi-inclusive $\PQc$ plus light-jet production at $\sqrt{s} = 14$ TeV (HL-LHC, left) and $100$ TeV (nominal FCC, right), and for $4 < \DY < 6$.
First ancillary panels below primary plots shows the ratio between $\LL$ or $\HENLOp$ and $\NLLp$.
Second ancillary panels exhibit the ratio between diquark and direct initial-scale fragmentation.
Uncertainty bands capture the net effect of MHOUs and phase-space multidimensional integration.}
\label{fig:I-k1b_YM}
\end{figure*}

All $\DY$-distributions exhibit a consistent trend: they decrease as $\DY$ increases. 
This behavior results from the interplay of two competing effects. 
On the one hand, the partonic hard factor grows with energy (and therefore with $\DY$), as predicted by high-energy resummation. 
On the other hand, this growth is notably suppressed due to collinear convolution with PDFs and FFs within the emission functions (refer to Eqs.~\eqref{LOPIF} and~\eqref{LOJIF}).

From a pure resummation perspective, two key observations can be made. 
First, the predictions exhibit good stability under MHOUs, with uncertainty bands staying within a relative size of $2.5$ in the HL-LHC scenario, and $1.5$ in the FCC one (see the first ancillary panels). 
Although these numbers may appear large if compared to fixed-order precision benchmarks, they indicate strong stability when viewed in the context of high-energy resummation phenomenology. 
Indeed, in standard BFKL studies of light-particle production, such as dijet or hadron-jet final states~\cite{Celiberto:2016hae,Celiberto:2017ptm,Bolognino:2018oth,Celiberto:2020wpk}, variations induced by NLL/NLO$^{(+)}$ corrections can often reach one or two orders of magnitude. 
In contrast, our results for exotic pentaquarks remain under control across the entire $\Delta Y$ spectrum.

Second, we observe that the $\NLLp$ bands are systematically narrower than their $\LL$ counterparts across the full $\Delta Y$ range, a typical outcome of reduced scale dependence at higher perturbative orders.
While full overlap between bands is not achieved, especially at intermediate and large $\Delta Y$, the reduced spread and partial approach observed in some bins remain suggestive of improved perturbative stability.
This observation aligns with previous findings on doubly and fully charmed tetraquarks~\cite{Celiberto:2023rzw,Celiberto:2024mab}, and supports the idea that forward semi-inclusive exotic production offers a promising environment to probe high-energy dynamics with controlled theoretical uncertainties.

From a fragmentation model perspective, we observe that the diquark channel consistently yields larger results compared to the direct channel, especially in the moderate to low $\DY$ region. 
Further examination of the second ancillary panels reveals that this trend becomes significantly more pronounced when full NLL corrections are applied.
This highlights a potentially significant interplay between heavy-flavor fragmentation and high-energy resummation, which warrants dedicated future studies.

The plots of Fig.~\ref{fig:I-k1b_YS} present predictions for the $|\vec{\kappa}_1|$-differential distribution in the production of pentacharm-plus-jet systems at 14~TeV HL-LHC (left) and 100~TeV FCC (right), with $\DY$ integrated over a ``lower'' bin, $2 < \DY < 4$.  
To facilitate comparisons with future experimental data, we adopt transverse-momentum bins of uniform size, each spanning 10~GeV.

Similarly organized, Fig.~\ref{fig:I-k1b_YM} shows distributions for the ``upper'' $\DY$-bin, $4 < \DY < 6$, which is contiguous to the previous range. 
In both figures, first ancillary panels below the main plots display the ratio of the $\LL$ or $\HENLOp$ cases (inclusively labeled as `${\rm [res]}$') to the $\NLLp$ predictions, whereas second ancillary panels depict the diquark-to-direct production ratio.

Overall, our transverse-momentum rates exhibit a steep decline with increasing $|\vec \kappa_1|$.  
The results demonstrate notable stability under energy-scale variations, with uncertainty bands in the first ancillary panels not exceeding a relative width of 20\%.

We note that the $\HENLOp$ to $\NLLp$ ratios generally stay below one, decreasing as $|\vec \kappa_1|$ increases.
In contrast, the $\LL$ to $\NLLp$ ratio exhibits an almost opposite behavior: it starts below one in the low-$|\vec \kappa_1|$ region but gradually increases as $|\vec \kappa_1|$ grows.
The explanation for these trends is complex, as they result from a combination of several interacting effects.

On the one hand, previous analyses on semi-hard processes have shown that the behavior of the NLL-resummed signal relative to its NLO high-energy background in singly differential transverse-momentum rates can vary depending on the process being considered.
For example, the $\HENLOp$ to $\NLLp$ ratio for the cascade-baryon plus jet channel consistently exceeds one, as shown in Fig.7 of Ref.~\cite{Celiberto:2022kxx}.
However, preliminary analyses of Higgs-plus-jet distributions, performed within a partially NLL-to-NLO matched accuracy, reveal a more complex pattern~\cite{Celiberto:2023dkr}.

Therefore, the observation that the $\HENLOp$ to $\NLLp$ ratio is less than one in the context of pentaquark-plus-jet tags appears to be a distinctive feature of this specific process, and is also shared by similar differential observables sensitive to the semi-inclusive emission of tetraquark-plus-jet systems~\cite{Celiberto:2023rzw,Celiberto:2024mab,Celiberto:2024beg}.
This suggests that the dynamics governing these emissions differs from those observed in other semi-hard reactions, underlining the unique interplay of high-energy and NLL resummation effects in these channels.

On the other hand, the behavior of the $\LL$/$\NLLp$ ratio is shaped by a complex interplay of different factors, particularly the nature of NLO corrections associated with the various emission functions.
For jet emissions, NLO corrections to the jet function are mostly negative~\cite{Ivanov:2012ms,Colferai:2015zfa}.

Conversely, NLO corrections from the perturbative $C_{gg}$ coefficient entering the hadron emission function are positive, while those from $C_{gq}$, $C_{qg}$, and $C_{qq}$ coefficients are negative~\cite{Ivanov:2012iv}.
This suggests that, depending on the transverse-momentum phase space, these corrections could partially cancel each other out, leading to varying effects on the $\LL$ to $\NLLp$ ratio.

For instance, in the case of $\Xi^-/\bar\Xi^+$ plus jet tags, the $\LL$ to $\NLLp$ ratio is found to exceed one~\cite{Celiberto:2022kxx}.
However, this behavior is less pronounced in other processes, such as the production of a doubly charmed tetraquark plus a jet~\cite{Celiberto:2023rzw}.
This discrepancy underscores how the unique dynamics of each process shape the balance between leading and next-to-leading logarithms, resulting in varying $\LL$ to $\NLLp$ ratios across different reactions.

This difference becomes particularly significant in the region of large transverse momenta $|\vec \kappa_1|$.
In such kinematic configurations, far from the symmetric regime where $|\vec \kappa_1| \simeq |\vec \kappa_2|$, the applicability of the high-energy resummation framework becomes more delicate.
In fact, these asymmetric configurations enhance the sensitivity to subleading corrections and to the interplay between resummation and fixed-order components, often resulting in broader uncertainty bands and more pronounced deviations between $\LL$ and $\NLLp$ predictions.
This behavior is not unexpected: it has already been observed in other forward observables with heavy-flavored final states~\cite{Bolognino:2021mrc,Celiberto:2024beg,Celiberto:2025dfe,Celiberto:2025ziy} and is a known limitation of resummed approaches when pushed outside their canonical domain of validity.

Finally, from the inspection of second ancillary panels of plots in Figs.~\ref{fig:I-k1b_YS} and~\ref{fig:I-k1b_YM}, it emerges that the diquark-like $\PQc$ initial-scale fragmentation brings to larger results with respect to the direct case.
In nearly all instances, the diquark predictions increase with $|\vec \kappa_1|$, and this increase is more significant when either truncated or fully resummed NLL corrections are included.

\section{Closing statements}
\label{sec:conclusions}

We investigated the leading-power fragmentation of fully charmed pentaquark states ($S$-wave $\PQc$ pentacharms) at next-generation hadron colliders.
To this extent, we derived a novel set of \emph{multimodal} collinear fragmentation functions, named {\tt PQ5Q1.0} determinations.
They are based on an enhanced calculation of the initial-scale input for the constituent (anti)charm fragmentation channel, making them well suited to describe the short-distance emission of either a compact multicharm state or a dicharm-charm-dicharm configuration.

To explore phenomenological implications, we made use of the numerical {\tt JETHAD} code and the {\tt (sym)JETHAD} symbolic-manipulation plugin~\cite{Celiberto:2020wpk,Celiberto:2022rfj,Celiberto:2023fzz,Celiberto:2024mrq,Celiberto:2024swu} to analyze semi-inclusive production rates for penta\-charm-plus-jet systems within the NLL/NLO$^+$ hybrid collinear and high-energy factorization formalism, and at center-of-mass energies ranging from 14~TeV~HL-LHC to 100~TeV~FCC.

The application of collinear fragmentation to describe the production of $\PQc$ particles at high transverse momentum effectively stabilized our high-energy resummation framework, mitigating potential instabilities caused by NLL corrections and nonresummed threshold logarithms.  
This resulting \emph{natural stability} validated the reliability and convergence of our formalism over a broad range of center-of-mass energies, spanning from the LHC to the FCC. 

To reach the precision level, we plan to enhance our $\NLLp$ factorization in a ``multilateral'' way, integrating additional resummation techniques.
The initial focus will be on establishing links with soft-gluon~\cite{Hatta:2020bgy} and jet-radius resummations~\cite{Dasgupta:2014yra,Banfi:2012jm,Banfi:2015pju,Liu:2017pbb}.  
Moreover, investigating potential synergies with ongoing studies on jet angularities~\cite{Caletti:2021oor} represents an exciting direction for future research.

The fragmentation production of heavy-flavored hadrons at leading power provides a unique and important venue where hadronic structure and precision QCD intersect. 
The presence of one or more heavy-quark species in the lowest Fock state has a twofold impact. 
On the one hand, it complicates the theoretical description of the initial-energy fragmentation compared to light hadrons. 
This necessitates a consistent modeling of the nonperturbative component, potentially capturing momentum and spin correlations among constituent partons. 

On the other hand, since heavy-quark masses are significantly larger than $\LQCD$, perturbative techniques are essential for accurately calculating the short-distance fragmentation component. 
Therefore, a precise description of heavy-flavor fragmentation requires leveraging \emph{the best of both worlds}, where hadron-structure explorations and precision QCD calculations coexist as fundamental ingredients.

The {\tt PQ5Q1.0} FFs offer valuable insights for exploratory studies on pentacharm production across a wide variety of processes, ranging from semi-inclusive emissions at next-generation hadron colliders to observations in lepton and lepton-hadron experiments.
Looking ahead, we aim to enhance our description of $\PQc$ fragmentation by incorporating a more robust uncertainty quantification, potentially linked to MHOU effects~\cite{Kassabov:2022orn,Harland-Lang:2018bxd,Ball:2021icz,NNPDF:2024dpb}.

Beyond their current formulation, the {\tt PQ5Q1.0} \emph{multimodal} functions lay the groundwork for a more comprehensive fragmentation framework in which multiple hadronic structures, such as compact multiquarks, diquark clusters, or loosely bound meson-baryon systems, can coherently contribute to the initial-scale modeling. 
As the parton undergoes energy loss and approaches the hadronization scale, quantum superpositions and dynamical transitions across different Fock components may become relevant, reflecting the internal complexity of exotic states. 
While the present implementation of the multimodal picture is still in its early stages, it may evolve into a powerful framework to access structural information from future high-precision data.

Moreover, calculating fragmentation contributions from other parton channels will serve two main purposes: $(i)$ enabling a comprehensive application of the {\HFNRevo} methodology~\cite{Celiberto:2024mex,Celiberto:2024bxu,Celiberto:2024rxa} by accounting for the semi-analytic decoupled evolution aspect of DGLAP ({\tt EDevo}, see Sec.~\ref{ssec:FFs-PQ5Q10} for more details), and $(ii)$ potentially distinguishing the production mechanisms of $\PQc$ and $\bPQc$ states.

A deeper understanding of the internal structure of hadrons will gradually develop as we advance our knowledge of the core dynamics responsible for quarkonium and exotic matter formation. 
This progress will be driven by data collected at the FCC~\cite{FCC:2018byv,FCC:2018evy,FCC:2018vvp,FCC:2018bvk} and other future colliders~\cite{Chapon:2020heu,Anchordoqui:2021ghd,AlexanderAryshev:2022pkx,InternationalMuonCollider:2024jyv,MuCoL:2024oxj,Black:2022cth,Accardi:2023chb}. 
It was recently pointed out that unresolved photoproductions of a $\Jpsi$ plus charmed-jet~\cite{Flore:2020jau} at the upcoming EIC~\cite{AbdulKhalek:2021gbh,Khalek:2022bzd,Abir:2023fpo} will provide a promising channel for measuring the intrinsic-charm~\cite{Brodsky:1980pb,Ball:2016neh,Hou:2017khm,Ball:2022qks,Guzzi:2022rca} valence PDFs in the proton~\cite{NNPDF:2023tyk}. 
This will serve as a bridge between intrinsic-charm phenomena inside hadrons and the physics of exotics~\cite{Vogt:2024fky}.

A connection between the intrinsic-charm puzzle effects and the existence of doubly-charmed pentaquark states was explored in Ref.~\cite{Mikhasenko:2012km}. 
Pentaquarks arising from intrinsic charm in $\Lambda_b$ baryon decays were studied in Ref.~\cite{Hsiao:2015nna}. 
Additionally, measurements of multi-$\Jpsi$ production rates made by the NA3 experiment in the 1980s, later corroborated by studies at the LHC and Tevatron, could provide insights supporting hypotheses involving pion double intrinsic-charm effects and tetraquark resonances~\cite{NA3:1982qlq}.

A promising avenue for future research involves exploring the connection between exotic heavy-flavor hadrons and the dead-cone effect in QCD, first theorized in~\cite{Dokshitzer:1991fd} and recently observed by the ALICE Collaboration~\cite{ALICE:2021aqk}. 
Fully heavy exotic observables, especially in jet-inclusive configurations, may offer new insights into heavy-quark dynamics and in-medium radiation suppression mechanisms.

Pentaquarks provide a unique testing ground for the nonperturbative regime of QCD, where confinement and hadronization dominate~\cite{Jaffe:2004ph,Esposito:2016noz}. Their study complements ongoing efforts to map the internal structure of hadrons, adding a novel dimension to the quark-gluon dynamics explored in baryons and mesons~\cite{Klempt:2009pi}.

\section*{Data availability}
\label{sec:data_availability}
\addcontentsline{toc}{section}{\nameref{sec:data_availability}}

The {\tt PQ5Q1.0} collinear FFs~\cite{Celiberto:2025_PQ5Q10} for $S$-wave pentacharms can be publicly accessed from the following url: \url{https://github.com/FGCeliberto/Collinear_FFs/}. 
They consist of a collinear set in {\tt LHAPDF} format. 
The central value (replica 0) represents the direct fragmentation channel, whereas replica 1 is for the diquark channel.

\section*{Acknowledgments}
\label{sec:acknowledgments}
\addcontentsline{toc}{section}{\nameref{sec:acknowledgments}}

We thank A.~Pilloni for insightful conversations on the physics of exotic hadrons and S.~M. Moosavi Nejad for a discussion on pentaquark fragmentation.
We sincerely thank A.~Papa for his critical reading of the manuscript, valuable suggestions, and encouragement.
We would like to express our gratitude to colleagues of the \textbf{Quarkonia As Tools} and \textbf{EXOTICO} Workshops for inspiring conversations and the warm atmosphere.
This work received support from the Atracci\'on de Talento Grant n. 2022-T1/TIC-24176 of the Comunidad Aut\'onoma de Madrid, Spain.

\appendix

\counterwithin*{equation}{section}
\renewcommand\theequation{\thesection\arabic{equation}}

\hypertarget{app:A}{
\section{Direct fragmentation coefficients}
}
\label{app:A}

The $\gamma_{P,\,{\rm [direct]}}^{(c)}(z; k)$ fragmentation coefficients entering Eq.~\eref{PQc_FF_initial-scale_Q_num_direct} and corresponding to the direct multicharm scenario are given by

\begin{subequations}
\allowdisplaybreaks 
\begin{align}
 \label{aA:eq:gamma_k}
 \nonumber
 \gamma_{P,\,{\rm [direct]}}^{(c)}(z; 0) &= (z+5)^2 (347638095 z^{18}-14333286144 z^{17}+303132501680 z^{16} \\ \nonumber
\,&-\, 
 2979357459428 z^{15}+20140712527168 z^{14}-101570323071060 z^{13} \\ \nonumber
\,&+\, 
 381426354056100 z^{12}-1070325481916500 z^{11}+2274630737903750 z^{10} \\
\,&-\, 
 3716109400562500 z^9+4726702575562500 z^8-4727934482187500 z^7 \\ \nonumber
\,&+\, 
 3748290871875000 z^6-2366985523437500 z^5+1187884179687500 z^4 \\ \nonumber
\,&-\, 
 464519335937500 z^3+134047607421875 z^2 \\ \nonumber
\,&-\, 
 25307617187500 z+2319335937500) \;, \\[0.50cm] \nonumber
 \gamma_{P,\,{\rm [direct]}}^{(c)}(z; 1) &= 1133852889 z^{18}-26979532892 z^{17}+241365645652 z^{16} \\ \nonumber
\,&-\, 
 406276817520 z^{15}-5088767528028 z^{14}+35085927681040 z^{13}
\\ \nonumber
\,&-\, 
 85928556891600 z^{12}-39913868698000 z^{11}+926214176418750 z^{10} \\
\,&-\, 
 3208940492025000 z^9+6562195688125000 z^8-9297361556250000 z^7 \\ \nonumber
\,&+\, 
 9712589920312500 z^6-7742747343750000 z^5+4790060781250000 z^4
\\ \nonumber
\,&-\, 
 2279664843750000 z^3+792703369140625 z^2 \\ \nonumber
\,&-\, 
 178178710937500 z+19165039062500 \;, \\[0.50cm]
 \nonumber
 \gamma_{P,\,{\rm [direct]}}^{(c)}(z; 2) &= 331172883 z^{16}-5540112194 z^{15}+24041536795 z^{14}+1829547432 z^{13} \\ \nonumber
\,&-\, 
 453100415317 z^{12}+1783755355930 z^{11}-1450496653425 z^{10} \\
\,&-\, 
 9989878632500 z^9+42225730018125 z^8-87820041068750 z^7 \\ \nonumber
\,&+\, 
 120577380890625 z^6-120190321250000 z^5+90757141015625 z^4 \\ \nonumber
\,&-\, 
 51981503906250 z^3+21462470703125 z^2 \\ \nonumber
\,&-\, 
 5640039062500 z+698242187500 \;, \\[0.50cm]
 \nonumber
 \gamma_{P,\,{\rm [direct]}}^{(c)}(z; 3) &= 161946833 z^{14}-1713576308 z^{13}+6158065036 z^{12}+2311752868 z^{11} \\ \nonumber
\,&-\, 
 64931596275 z^{10}+119000823740 z^9+171719850500 z^8 \\
\,&-\, 
 1135906421000 z^7+2497182556875 z^6-3447541987500 z^5 \\ \nonumber
\,&+\, 
 3411658187500 z^4-2484448437500 z^3+1270333984375 z^2 \\ \nonumber
\,&-\, 
 402773437500 z+58789062500 \;, \\[0.50cm]
 \nonumber
 \gamma_{P,\,{\rm [direct]}}^{(c)}(z; 4) &= 59591417 z^{12}-431073302 z^{11}+687052153 z^{10}+1761733340 z^9 \\
\,&-\, 
 6636037130 z^8+1595946780 z^7+28907583850 z^6-80523124000 z^5 \\ \nonumber
\,&+\, 
 124845798125 z^4-129207593750 z^3+87968828125 z^2 \\ \nonumber
\,&-\, 
 35317187500 z+6289062500 \;, \\[0.50cm]
 \nonumber
 \gamma_{P,\,{\rm [direct]}}^{(c)}(z; 5) &= 5173335 z^{10}-23475428 z^9+16082740 z^8+88328984 z^7 \\
\,&-\, 
 200793674 z^6-4722280 z^5+759004600 z^4-1640425000 z^3 \\ \nonumber
\,&+\, 
 1744896875 z^2-964862500 z+220937500 \;, \\[0.50cm]
 \gamma_{P,\,{\rm [direct]}}^{(c)}(z; 6) &= 121839 z^8-324922 z^7-210833 z^6+1991920 z^5-2327663 z^4 \\ \nonumber
\,&-\, 
 2475170 z^3+8563925 z^2-7864500 z+2537500 \;, \\[0.50cm]
 \gamma_{P,\,{\rm [direct]}}^{(c)}(z; 7) &= 3277 z^6-2756 z^5-17820 z^4+30188 z^3 \\ \nonumber
\,&-\, 
 17647 z^2-65740 z+36500 \;, \\[0.50cm]
 \gamma_{P,\,{\rm [direct]}}^{(c)}(z; 8) &= 183 z^4+262 z^3-769 z^2-532 z+1180 \;, \\[0.50cm]
 \gamma_{P,\,{\rm [direct]}}^{(c)}(z; 9) &= (z+2)^2 \;.
\end{align}
\end{subequations}

\clearpage
\bibliographystyle{apsrev}
\bibliography{bibliography}

\end{document}